# Deciphering contact interactions and exploration strategies underlying tactile perception of material softness

A

Dissertation

Presented to

the faculty of the School of Engineering and Applied Science

University of Virginia

in partial fulfillment

of the requirements for the degree

Doctor of Philosophy

by

Chang Xu

December 2021

# APPROVAL SHEET

This
Dissertation
is submitted in partial fulfillment of the requirements
for the degree of
Doctor of Philosophy

Author: Chang Xu

This Dissertation has been read and approved by the examing committee:

Advisor: Gregory J. Gerling

Advisor:

Committee Member: Sara Lu Riggs

Committee Member: Afsaneh Doryab

Committee Member: Seongkook Heo

Committee Member: Hong Z. Tan

Committee Member:

Committee Member:

Accepted for the School of Engineering and Applied Science:

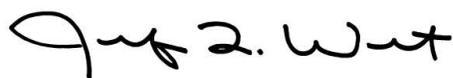

Jennifer L. West, School of Engineering and Applied Science

December 2021

# Abstract

Our sense of touch is essential and permeates in interactions involving natural explorations and affective communications. For instance, we routinely judge the ripeness of fruit at the grocery store, caress the arm of a spouse to offer comfort, and stroke textiles to gauge their softness. Meanwhile, interactive displays that provide tactile feedback are becoming normal and ubiquitous in our daily lives, and are extending rich and immersive interactions into augmented and virtual reality. To replicate touch sensation and make virtual objects feel tangible, such feedback will need to relay a sense of compliance, or "softness", one of the key dimensions underlying haptic perception. As our understanding of softness perception remains incomplete, this study seeks to understand exploratory strategies and perceptual cues that may optimally encode material softness. Specifically, we employ methods of computational finite element modeling, biomechanical experimentation, psychophysical evaluation, and data-driven analysis. First, we characterize the functional roles of physical contact cues, by studying a tactile illusion phenomenon where small-compliant and large-stiff spheres are naturally indistinguishable. In modulating contact interactions between the finger pad and stimuli, we found that pressing an object into the finger does not fully reveal its softness, but pressing actively does. Thus, our percept of softness is a product of both sensation and volition and depends upon both tactile afferents in the skin and musculoskeletal proprioception. Second, in exploring both engineered and ecological soft objects, we investigate how cues are optimally evoked and integrated under one's active control. By varying exploration time, we observe that exploratory strategies are finely tuned to elicit efficient contact force and finger movements for superior performance. Third, considering inherent differences and constraints among individuals' skin mechanics, we investigate individual differences in eliciting tactile cues, exploratory strategies, and thus, perceptual sensitivity. We characterized the skin material properties of individuals' finger pads and evaluated their contact interactions in performing discriminative touch. The results indicate that an individual's tactile acuity is constrained by their skin softness, but could be improved under volitional control of their exploratory movements. Overall, this work may aid in engineering the next-generation wearable haptic displays, which must be more tangible, compatible, and perceptually naturalistic.

# Contents

# 1   Overview of Aims

Our sense of touch helps us encounter the richness of our natural world. Across a myriad of contexts and repetitions, we have learned to deploy touch contacts in daily activities. When we grasp a soft, deformable object, our somatosensory system transforms compliance, an intrinsic material property, into percepts of 'softness' or 'hardness.' The task of identifying optimal physical cues and perceptual strategies that underlie our softness perception remains relevant and incomplete. Meanwhile, to design consumer displays that render touch sensations, we need to tease apart mechanical signals from contact, their transformations from physical to perceptual space, and whether they are perceptually integral or separable. Furthermore, most studies consider only a cohort of individuals, though aggregated empirical results, where diverse physical and behavioral interactions among individuals are yet to be investigated.

The objective of this study is to decipher optimal contact interactions and strategies that underlie softness perception within and among individuals. The central hypothesis is that the optimal exploratory and perceptual strategies govern the modulation of spatiotemporal signals from skin deformation and finger proprioception, and thus, the encoding of our softness perception. We seek to address this by employing the combination of computational finite element modeling of skin materials and contact mechanics, along with biomechanical measurements of physical contact and exploratory movements, psychophysical evaluations of signal detection sensitivity, and statistical and data-driven analyses of time-series correlations between skin mechanics, physical contact, and perceptual responses. To this end, we have assembled mechanical and electrical test apparatus, constructed and calibrated test stimuli, designed and conducted human-subjects experiments.

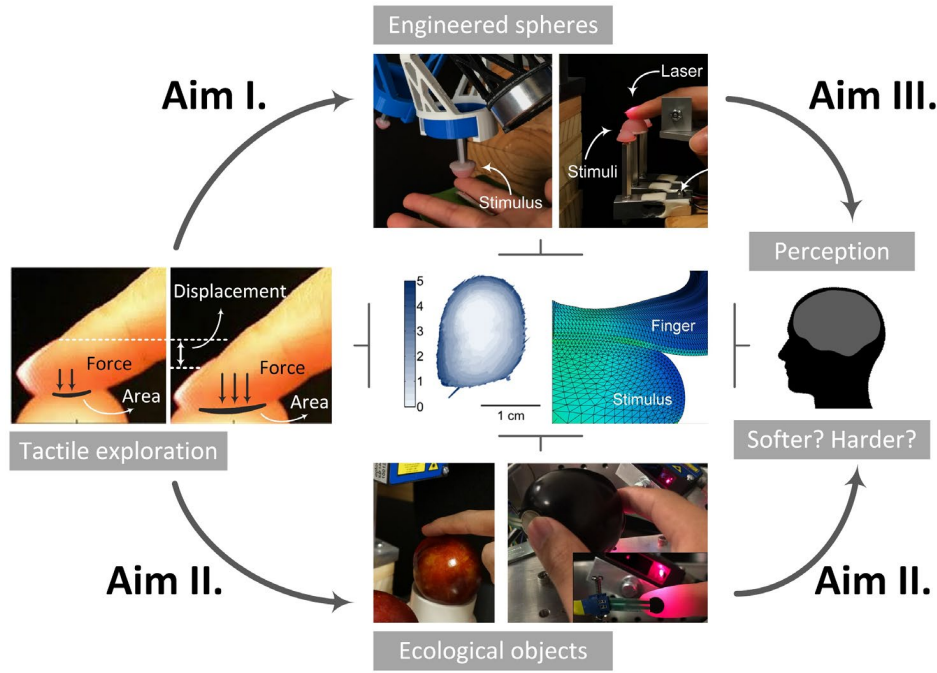

**Figure 1.1 Scope of this dissertation.** Aim I characterize the functional roles of cutaneous and kinesthetic cues in encoding material softness. Aim II investigate optimal movement strategies and perceptual rules for tactile exploration of both engineered and ecological soft objects. Aim III decipher individual differences in skin mechanics, exploratory strategies, and thus, softness perception.

**Table 1.1 Summary on the Scope of the Dissertation.** Experimental conditions, findings, and publications.

| Section | Experimental conditions | Findings | Publications |
|---|---|---|---|
| Aim I. | 1. Passive touch w/ controlled touch force<br>2. Active touch w/ behaviorally controlled force | Softness perception is a product of cutaneous and proprioceptive cues. | HFES 2018, PLOS Comput. Biol. 2021 |
| 4.3.1 – 4.3.4 | 1. Active touch w/ behaviorally controlled force<br>2. Fully active touch | Tactile exploration strategies elicit virtual stiffness cues.<br>Engineered stimuli approximate ecological objects. | WHC WIP 2019, WHC 2019, IEEE Trans. Haptics 2020 |
| 4.3.5 & 4.3.6 | Fully active touch | Perceptual strategies are modeled by temporal tactile cues | HAPTICS 2020 |
| 4.3.7 & 4.3.8 | 1. Active touch w/ controlled exploration time<br>2. Fully active touch | Perceptual strategies are fine-tuned to elicit optimal cues in tactile decision making | J. R. Soc. Interface (in writing) 2022 |
| Aim III. | 1. Passive touch w/ controlled touch force<br>2. Active touch w/ behaviorally controlled force | Individual softness perception is constrained by skin mechanics but improved under active control | WHC 2021 |

**Aim I. Characterize physical contact cues by an elasticity-curvature tactile illusion phenomenon.** To decipher the precise sets of tactile cues that optimally drive our perception, we investigated a tactile illusion phenomenon associated with softness, specifically, in exploring spherical stimuli with covaried elasticity and curvature. In particular, finite element modeling of the distal finger pad was used to develop elasticity-curvature combinations that afford non-differentiable cutaneous contact cues. Then, experimental test platforms were assembled to investigate biomechanical contact mechanisms that underlie this potential illusory phenomenon both in passive and active touch interactions. These biomechanical attributes were also captured amidst human-subjects psychophysical evaluations in tasks of discriminating softness. Through this combination of solid mechanics modeling, biomechanical contact measurement, and psychophysical evaluation, we dissociated and clarified relative contributions from cutaneous and proprioceptive cues in encoding our softness perception.

**Aim II. Decipher optimal strategies and perceptual modeling in tactile explorations among populations.** We characterized the optimal strategies for active control of exploratory movements that can more efficiently link with and elicit perceptual cues, especially those mediated by interactions with naturalistic objects where our somatosensory system has been finely tuned. In particular, human-subjects experiments were conducted for the active exploration of soft objects, both with engineered and ecological stimuli. Exploratory strategies of active control of tactile movements were investigated for different discrimination tasks, along with psychophysical evaluations. Furthermore, two potential perceptual strategies were modeled for the decision-making process in discriminating softness. The utility and employment of these perceptual strategies were further investigated by applying perceptual models with measured spatiotemporal tactile cues.

**Aim III. Individual difference in skin mechanics, exploratory strategies, and perceptual sensitivity.** We seek to understand the extent of the impact of skin mechanics and perceptual strategies in modulating individual perceptual performance. In particular, the skin mechanics was quantified by performing a standard uniaxial compression test on individuals' finger pads. Material properties were then characterized by fitting the force-displacement and stress-strain curves derived from the compression. Next, human-subjects experiments were conducted for active exploration of soft objects, where biomechanical measurement on contact cues and psychophysical evaluations were combined. The quantified tactile acuity for individuals was correlated with corresponding skin material properties and exploratory movements, thus revealing how individual differences in tactile perception could be attributed to distinct skin mechanics and perceptual strategies.

## 2 Background

**Tactile rendering displays and spatiotemporal tactile cues.** At present, several groups are working to build tactile displays to render a sense of compliance, or "softness" [1]–[4]. They have used mechanisms of tilting plates, concentric rings, particle jamming, and fabric stretch. The task of aligning actuation mechanisms with the most efficient, salient, and naturalistic perceptual cues remains relevant and timely. Many efforts have focused on cues of contact area, skin deformation, and kinesthetic inputs of force and joint angles [5]–[7]. However, replicating these stationary cues of contact area as a function of force does not afford the same perceptual acuity as naturalistic objects [5], [7]–[9]. For this reason, besides characterizing terminal contact force and gross contact area, physical cues of a time-dependent nature may also underlie our percepts of softness. Indeed, information in the rate of change of skin deformation, penetration of contact surface, the evolution of interior stress distribution, and dynamics of finger joints are suggested to improve efficiency and fidelity in conveying softness. In fact, for the case of differentiating stimulus curvature and angle, the relative timing of just the first spikes elicited in tactile afferents reliably conveys such spatial information [10]. At the behavioral level, the availability of force-rate cues can make the compliant objects more readily discriminable, by reducing the amount of skin deformation [8].

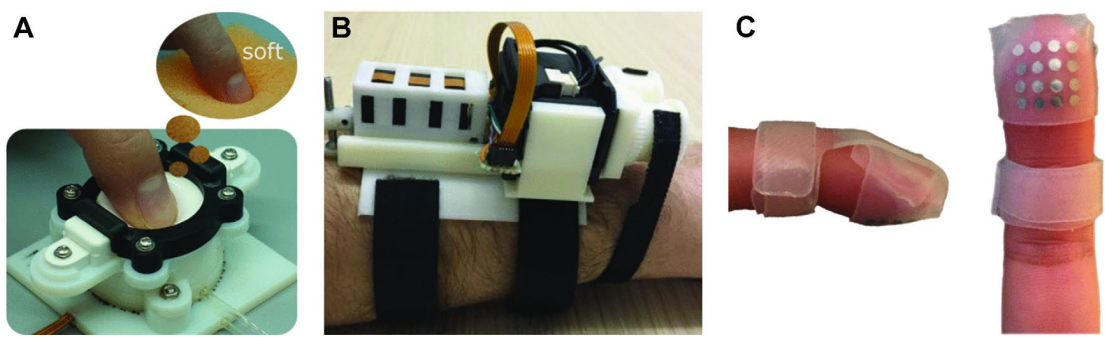

**Figure 2.1 Tactile displays adopted from the most recent publications.** (A) A pneumatic concave deformable device for hardness perception [11]. (B) A wearable haptic device rendered by squeezing and stroking [12]. (C) A soft wearable device using lateral skin stretch [13].

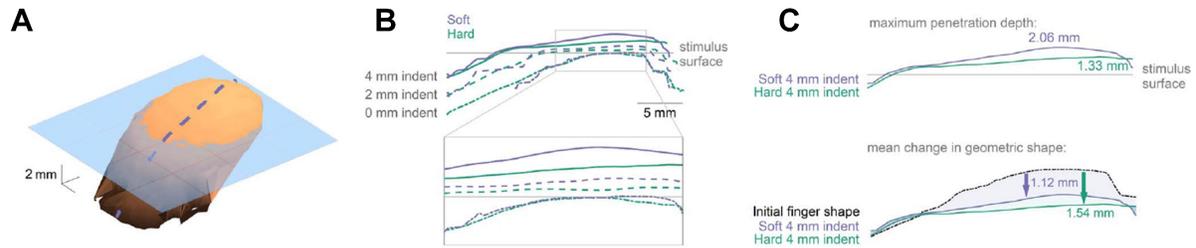

**Figure 2.2 Characterizing the finger pad surface as it contacts and deforms compliant stimuli.** Adopted from [14]. (A) Spatiotemporal changes in the contact cross-section are detailed in (B) as obtained from 3D surface reconstruction of finger pad-stimulus contact. (C) Temporal cues of penetration depth and surface deformation may differentiate stimulus softness.

Likewise, hand-held, probe-based studies have suggested that judgments are based upon force-rate and velocity information, as opposed to steady relationships of force and displacement [15].

**Psychophysics of softness exploration and perceptual strategies.** Human ability to differentiate and recognize compliant materials has been investigated in psychophysical studies to understand underlying sensorimotor mechanisms and to begin to build tactile and proprioceptive user interfaces. An early study found that human could differentiate objects' softness if their compliances differed by ~13% [16]. Another set of softness discrimination tasks was conducted under passive touch, active touch, and proprioceptive feedback only [7]. They found that independent of proprioception, surface compliances can be accurately differentiated using passive tactile information only. These studies were further extended and findings indicated that participants could discriminate compliances with Weber fractions of ~15% and tactile cues related to spatiotemporal deformation of skin surface might be vital [5]. Among these studies, particular psychophysical methods were determined and used in part by the specific research question. It is clear that reliable and valid psychophysical data can be derived if the conditions of stimulus presentations are finely controlled and the tasks assigned to participants are precisely specified [19]. Furthermore, as for perceptual strategies in exploring soft objects, studies indicated that the duration and sequence of exploration procedures also affect compliance discrimination. Discrimination with rubber stimuli showed that a longer duration of exploration with slower imposed force may result in a compliance judged softer [20]. When exploring objects sequentially, relevant sensory inputs are progressively gathered, retained, and integrated over time. Therefore, subsequent percepts might encode the fading representation of previous explorations and not all information contributes equally to the final percepts [21], [22]. For instance, a higher magnitude of touch force will be used in discriminating the softness of two objects, when these two are expected to be harder than two other softer objects [23]–[26].

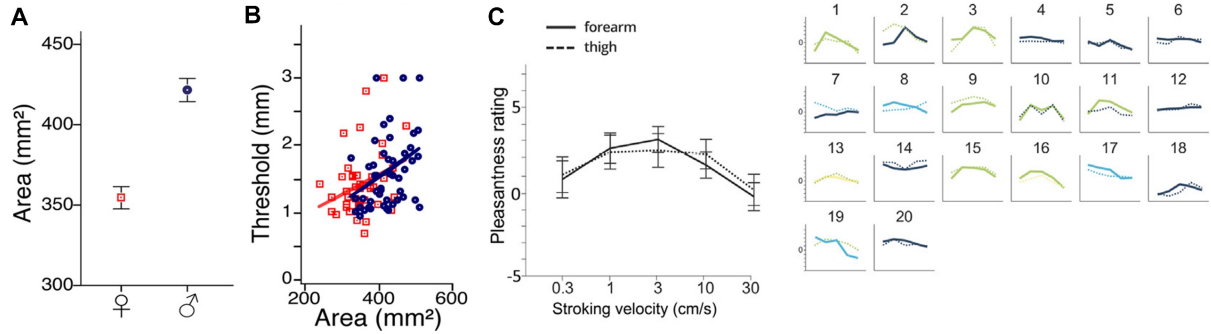

**Figure 2.3 Individual differences in skin contact mechanics and tactile perception.** Distinct surface contact area (A) and detection threshold in grating orientation task (B) by the index finger distal phalanx, adopted from [17]. (C) Ratings of pleasant tactile perception over the stroking touch velocity among and within individuals. Variations could be observed between individuals' performance. Adopted from [18].

**Individual differences in skin mechanics and tactile sensitivity.** As the primary interface for tactile explorations, the material properties of an individual's finger pad skin are suggested to modulate the mechanical transformation of conveying stimulus from the surface to interior layers where tactile end-organs reside [27]–[29], thus, may directly affect their perceptual judgments [30]–[32]. As a nonlinear, hyperelastic material, the thickness, stiffness, and elasticity of skin change dramatically over the lifespan, across skin sites, and among individuals [33], [34]. For instance, increased epidermal thickness separates mechanoreceptors away from the surface pressure, thus, attenuates afferent firing rates at the perceptual threshold [28], [35]. Meanwhile, the elasticity directly impacts the magnitude of skin deformation, thus, affecting interior stress conveyed to both Merkel cell-neurite complexes and Meissner corpuscles, which underlie individuals' touch perception [29], [36]. In particular, the increased skin elasticity, or "hardness", could render a lower perceptual sensitivity to surface pressure stimulus [28], [37]. Similar to elasticity, individuals with more compliant finger pad skin exhibit substantially lower perceptual thresholds in grating orientation discrimination, especially for younger individuals [30]. Furthermore, as compared with hairy skin sites, finger pad skin exhibits a higher density of myelinated afferents and is more receptive to touch [34]. Moreover, there is gathering evidence showing that individuals' tactile acuity differs between fingertip size and gender. Specifically, with more densely distributed Meissner corpuscles, individuals with smaller finger sizes are more sensitive to light touch and vibrations [17], [36]. Similarly, the higher density of Merkel receptors at the finger pad renders superior tactile spatial acuity of women compared to men [17]. Therefore, distinct skin material properties among individuals might affect tactile acuity by modulating specific mechanical responses and mediating tactile afferents' firing properties.

# 3 Aim I. Characterize Physical Contact Cues by An Elasticity-Curvature Tactile Illusion Phenomenon

## 3.1 Introduction

We integrate a multimodal array of sensorimotor inputs in the everyday perception of our natural environment. Along with vision and audition, our sense of touch is essential in interactions involving dexterous manipulation, affective connections, and naturalistic exploration [38]–[41]. For example, we routinely judge the ripeness of fruit at the grocery store, caress the arm of a spouse to offer comfort, and stroke textiles to gauge their roughness and softness [9], [42], [43]. We seamlessly do so by recruiting sensorimotor inputs, fine-tuning motor control strategies, comparing current percepts to our prior expectations, and updating internal representations [44].

Historically, tactile illusions have revealed inherent interdependencies of our sensorimotor and perceptual systems. Among the many illusions identified [45], [46], the "size-weight" illusion is particularly well-known. It involves picking up two objects of identical mass but of varied volume, and indicates that the smaller object is generally perceived as heavier [47]. The size-weight illusion reveals a separation of our sensorimotor and perceptual systems in estimating an object's mass. In particular, while our sensorimotor system adapts to the mismatch between the predicted and actual signals to dynamically adjust our exploratory motions, our perceptual system recalibrates the size-weight relationship more gradually on a different time scale [45], [48], [49]. Another intriguing illusion regards our perception of curvature where a physically flat surface is manually explored along a lateral direction. Depending on the relative inward/outward motions of the surface and the observer's finger, the flat surface can be perceived as being convex or concave [45], [50]. The curvature illusion reveals a poor spatial constancy of our somatosensory system, driven by a dissociation between cutaneous and proprioceptive inputs [41]. A further illusion, by analogy with the Aubert-Fleischl phenomenon in vision, indicates a possible misperception in speed by touch [51]. In particular, observers are asked to estimate the speed of a moving belt stimulus. Compared with tracking the stimulus with a guided arm movement, where the finger is moving along with the belt's motion (i.e., proprioception is available), observers can overestimate the stimulus speed by touching the stimulus with a stationary hand (i.e., tactile cues only). These and other illusions shed light upon interdependencies of our sensorimotor and perceptual systems, i.e., processing mechanisms for the perception of object properties, e.g., size, orientation, and movement, are distinct from those underlying the mediation of those properties in sensorimotor control [52]–[55]. Furthermore, tactile illusions can serve as a tool in engineering applications where human perception could be manipulated, e.g., the "size-weight" illusion could be exploited to create particular stimuli in virtual reality whose physical properties may be perceived as changing during interactions. Meanwhile, illusions have also been considered as a metric to evaluate virtual environments by correlating the perceived realism with the illusion strength [45].

Among the many dimensions of touch, which include surface roughness, stickiness, geometry, and others, our perception of softness is central to everyday life [39]. Our understanding of tactile compliance, a key dimension of an object's "softness," remains incomplete. This percept is informed by some combination of cutaneous inputs from mechanosensitive afferents signaling skin deformation and proprioceptive inputs signaling body movements. Efforts to define the precise cues within skin deformation and body movements have focused on contact area [56]–[60], spatiotemporal deformation of the skin's surface [5], [8], [29], and kinesthetic inputs of displacement, force, and joint angle [6], [7], [23], [61]. Such an array of sensory contact inputs, mediated by independent cortical mechanisms, are recruited and integrated in the primary somatosensory cortex, and form the perceptual basis from which compliances are recognized and discriminated [62]. That being said, it yet remains unclear which exploratory movements could elicit those perceptual cues that most optimally encode material softness.

Here, we investigate a tactile illusion associated with softness perception, specifically, in exploring spherical stimuli with covaried elasticity and curvature. These physical attributes are routinely encountered, such as in judging the ripeness of spherical fruit. The illusion phenomenon is observed only in passive touch when the finger is stationary and only non-distinct cutaneous cues of interior stress and gross contact areas are available for perception. The spheres, however, become readily discriminable when explored volitionally in active touch where finger proprioception is involved. The spheres therefore naturally dissociate relative contributions from cutaneous and proprioceptive cues in encoding softness, and shed light on how we volitionally explore compliant objects in everyday life.

## 3.2  Results

We introduce a novel elasticity-curvature illusion where small-compliant and large-stiff spheres are perceived as indiscriminable in passive touch. These spheres are explored using single, bare finger touch. Our methodological paradigm is unique in that computational models of the skin's mechanics define the stimulus attributes prior to evaluation in human-subjects experiments. In particular, finite element models of the distal finger pad are used to develop elasticity-curvature combinations that afford non-differentiable cutaneous cues. Then, investigation of the mechanisms that underlie this potential illusory experience is done empirically with human-subjects via measurements of biomechanical interactions and evaluations of psychophysical responses. The results suggest that we use a force-controlled movement strategy to optimally evoke cutaneous and proprioceptive cues in discriminating softness.

First, the skin mechanics of the index finger are modeled with finite elements in simulated interactions with spherical stimuli. The models predict that small-compliant (10 kPa–4 mm) and large-stiff (90 kPa–8 mm) spheres will generate nearly identical cutaneous contact cues, which may render them indiscriminable in passive touch. In contrast, when the models simulate conditions of active touch, the resultant fingertip displacements with controlled force loads are found to be distinct, which may render them discriminable.

Next, driven by the model predictions, a series of biomechanical and psychophysical evaluations are conducted with human participants. The results reveal that these spheres are indeed indiscriminable when explored in passive touch with only cutaneous cues available. However, this phenomenon vanishes when cues akin to proprioception are systemically augmented by a participant's use of a force control movement strategy.

### 3.2.1 Experiment 1: Computational Modeling of the Elasticity-Curvature Illusion

Finite element analysis was performed to simulate the skin mechanics of the bare finger interacting with compliant stimuli. The material properties of the model were first fitted to known experimental data. Then, numerical simulations were conducted with spherical stimuli of covaried radius (4, 6, and 8 mm) and elasticity (10, 50, and 90 kPa). In two interaction cases, the fingertip was moved and constrained to simulate active and passive touch, respectively. To help quantify the discriminability of the spheres, response variables were derived from the stress distributions at the epidermal-dermal interface, where Merkel cell end-organs of slowly adapting type I afferents and Meissner corpuscles of rapidly adapting afferents reside, as well as the required fingertip displacements to a designated force. The former was deemed as the cutaneous cue [10], [29], [63], [64]. The latter was associated with proprioception where displacement approximates the change in muscle length and force tied to muscle tension [6], [64]–[67].

In the simulation of passive touch where only cutaneous cues are available, the compliant spheres deformed the surface of the skin distinctly for each combination of elasticity and radius (Fig 3.1). Spatial distributions of stress for both the finger pad and spheres were simulated to a steady-state load of 2 N.

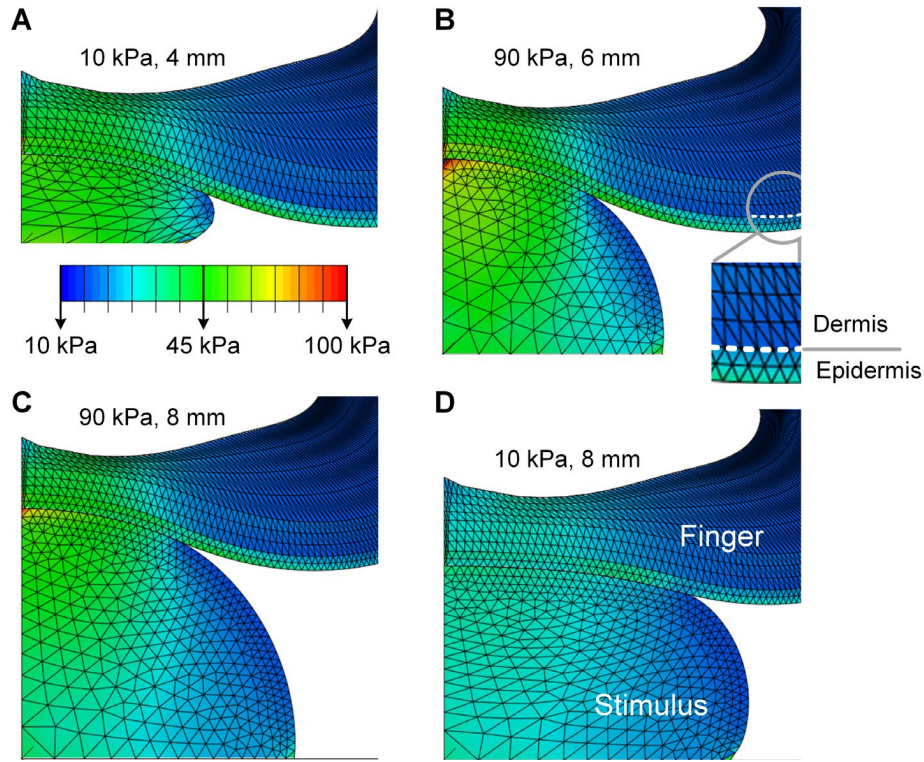

**Figure 3.1 Computational modeling of contact mechanics with compliant spheres.** Spatial distributions of stress are simulated at a load of 2 N for contact with spheres of (A)10 kPa-4 mm, (B) 90 kPa-6 mm, (C) 90 kPa-8 mm, and (D) 10 kPa-8 mm respectively. The epidermal-dermal interface was indicated in (B) and was consistently modeled for all simulation conditions. Although the deformation of the spherical stimuli differs greatly from (A) to (C), the resultant stress distributions and surface deflection at the finger pad are nearly identical.

For all the nine spheres simulated, either an increase of the spherical radius or a decrease of the elasticity decreases the concentration of stress quantities at contact locations, with the lowest stress concentration for the10 kPa-8 mm sphere and the highest for the 90 kPa-4 mm sphere (detailed in Fig 3.2). Note that the 10 kPa-8 mm sphere was taken as the comparison case in the following analyses.

However, for certain elasticity-radius combinations, changes in the spheres' radii counteracted the changes in their elasticity, resulting in nearly identical stress distributions for cutaneous contact. Although the deformation of the stimuli differed vastly between the 10 kPa-4 mm (Fig 3.1A) and 90 kPa-8 mm spheres (Fig 3.1C), the surface deformation and stress distributions of the finger pad were quite similar. Specifically, stress distributions at the epidermal-dermal interface were nearly identical between the small-compliant (10 kPa-4 mm) and large-stiff (90 kPa-8 mm) spheres across all levels of load (Fig 3.3A). A similar case was demonstrated for the 10 kPa-4 mm and 90 kPa-6 mm spheres where the stress curves fairly well overlapped (Fig 3.3B), as compared to the distinct stimulus (Fig 3.3C).

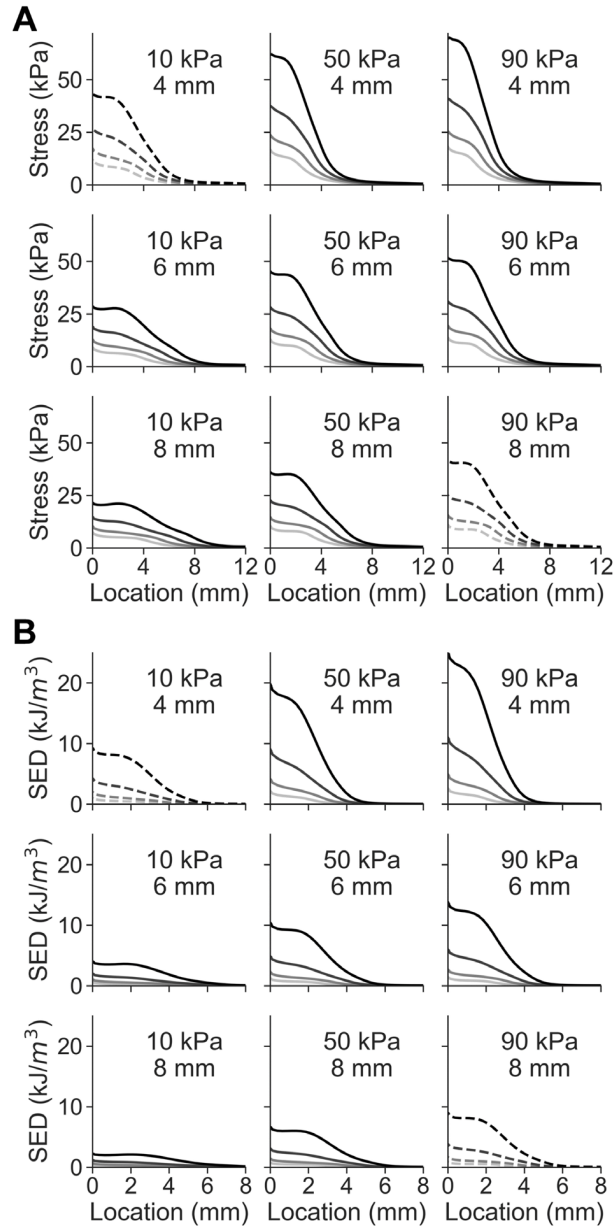

**Figure 3.2 Simulated spatial distributions of cutaneous cues.** (A) Spatial distributions of stress at contact locations for all nine spherical stimuli. (B) Spatial distributions of SED at the same contact locations for all spheres varying in radii and elasticity.

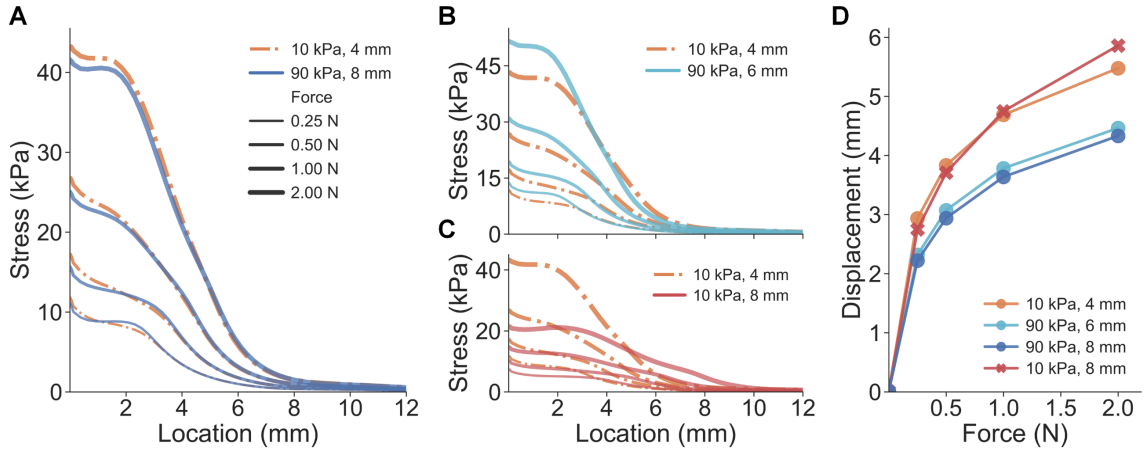

**Figure 3.3 Results of experiment 1: cues of cutaneous contact and proprioception**. (A) For the small-compliant (10 kPa-4 mm) and large-stiff (90 kPa-8 mm) spheres, stress distributions at the epidermal-dermal interface are nearly identical across all force loads. (B) Curves of stress distributions fairly well overlap for the 10 kPa-4 mm and 90 kPa-6 mm spheres. (C) Distinct stress distributions were obtained for spheres with the same elasticity but varying radii. (D) Proprioceptive cues of finger displacement are simulated for all force loads.

In addition to spatial distributions of stress, other response variables were also evaluated. The strain energy density (SED) at the epidermal-dermal interface and the deflection of the skin's surface were calculated and analyzed. Besides the stress/strain distributions, deflection of the skin surface – quantified by displacements at the node of the epidermis surface - is often considered as a cutaneous cue informing the change of contact area [56], [58]. Similar to the results in Fig 3.3, SED distributions and skin surface deflection from the three spheres (10 kPa-4 mm, 90 kPa-6 mm, and 90 kPa-8 mm) were nearly inseparable, which were predicted to generate indiscriminable contact area cues upon contact (Fig 3.4, detailed in Figs 3.2 and 3.5). In addition, mean values of cutaneous responses over the contact region were also similar between the three spheres (Fig 3.5). These results demonstrate that small-compliant and large-stiff stimuli can generate nearly identical cutaneous contact cues, therefore, non-informative for discriminating compliances whereas proprioceptive cues may be useful. It indicates that in passive touch where only cutaneous cues are perceptible, one might be unable to differentiate the aforementioned spheres. Therefore, these three stimuli (10 kPa-4 mm, 90 kPa-6 mm, and 90 kPa-8 mm) were denoted as the "illusion case spheres."

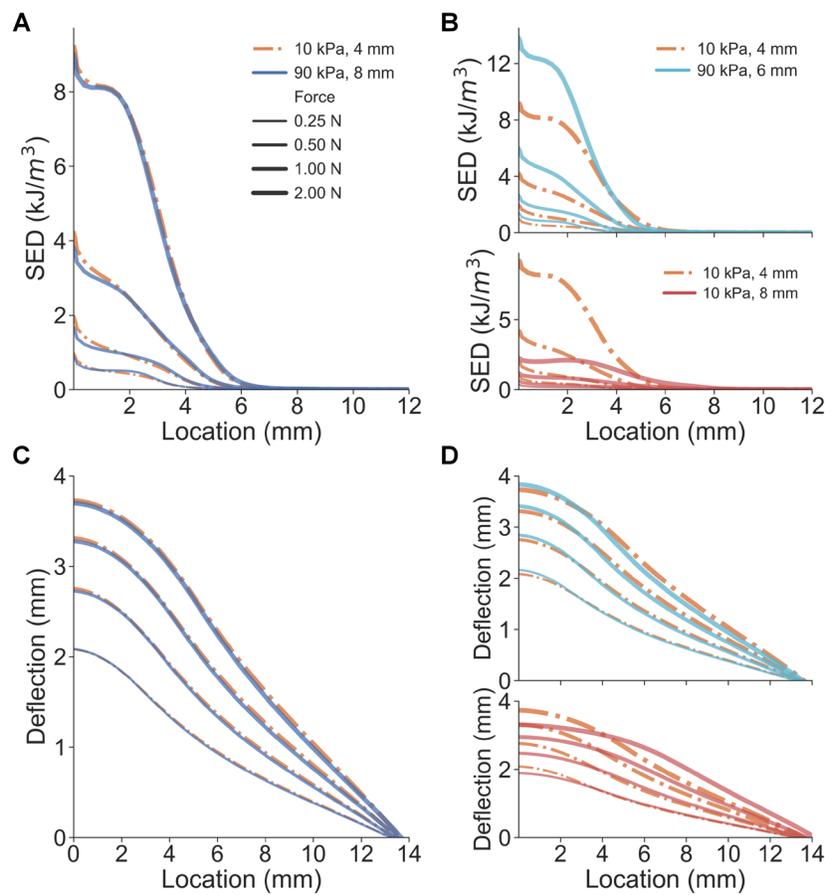

**Figure 3.4 Comparison of cutaneous cues between illusion and distinct spheres.** (A) Spatial distributions of SED are nearly identical for the small-compliant and large stiff spheres. (B) As opposed to the 10 kPa-8 mm sphere, SED distributions fairly well overlap between the 10 kPa-4 mm and 90 kPa-6 mm spheres. (C) Non-distinct surface deflection cues are obtained. (D) Consistent with SED distributions, surface deflections overlap for the 10 kPa-4 mm and 90 kPa-6 mm spheres.

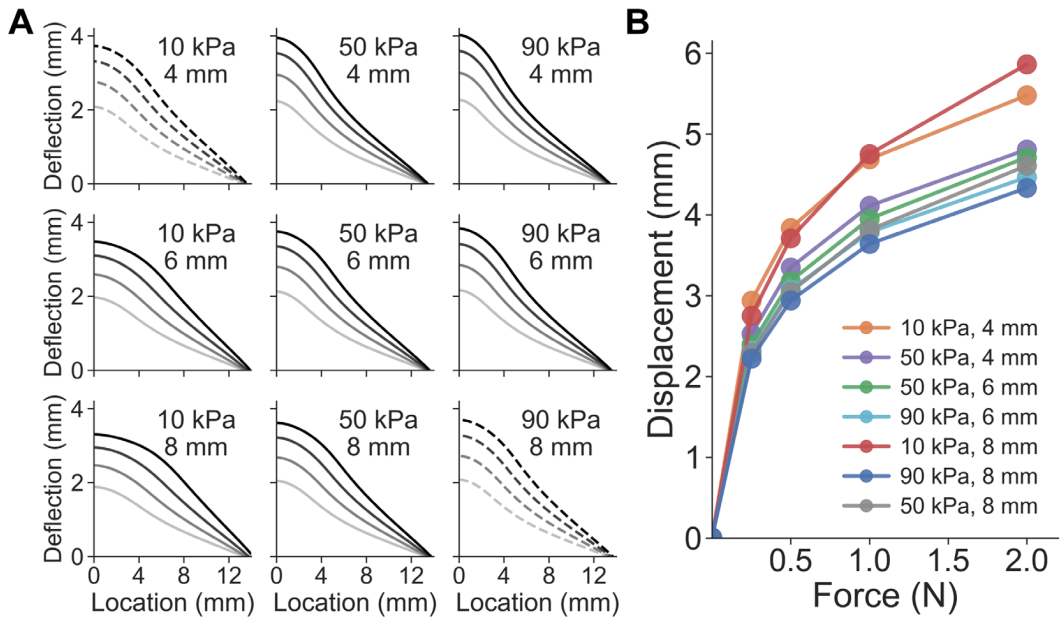

**Figure 3.5 Cues of the surface deflection and finger displacement.** (A) Simulated surface deflection of nodes at the surface of the finger pad model for all nine spheres. (B) Force-displacement relationships of the fingertip simulated for elasticity-radius combinations.

In the simulation of active touch, where both cutaneous and proprioceptive cues are available, an increase in either the radius or elasticity decreases the fingertip displacement given the same load (Fig 3.5). Specifically, the force-displacement curve of the 10 kPa-4 mm sphere was clearly separable from the 90 kPa-8 mm sphere (Fig 3.3D). Additionally, spheres of the same elasticity yielded overlapping force-displacement curves, as opposed to spheres of different elasticity. These results demonstrate that distinct proprioceptive cues tied to fingertip displacement differ given the indentation of the small-compliant compared to the large-stiff spheres. In active touch, where cues tied to fingertip displacement are utilized, one might be able to perceptually discriminate those illusion case spheres (10 kPa-4 mm, 90 kPa-6 mm, and 90 kPa-8 mm) amidst non-differentiable cutaneous contact cues.

Besides analyzing response variables only at the steady-state, the stimulus-ramp phase was further simulated to evaluate how contact mechanics would derive responses during the dynamic contact (detailed in [68]). Overall, the illusion case spheres could still afford nearly identical cutaneous responses during the stimulus ramp. The rate of change in stress distributions, SED, and surface deflection cues consistently overlap. This indicates that, throughout contact time-course done *in silico*, similar afferent responses from both slowly and rapidly adapting mechanoreceptors might be elicited among the illusion case spheres, and thus, may render an illusory experience in discriminating their compliances.

### 3.2.2  Experiment 2: Biomechanical Measurement of Cutaneous Contact

Derived from the computational analysis in Experiment 1, we hypothesized that similar cutaneous contact cues might be observed among the illusion case spheres. To validate this prediction, we conducted biomechanical measurement experiments with human-subjects.

In particular, through a series of biomechanical measurements, the contact area between the finger pad and stimulus was quantified to determine if the illusion case spheres would generate similar cutaneous contact profiles. The contact area was measured directly, using an ink-based procedure [8]. The measured contact area is commensurate with the cutaneous cues predicted in the finite element simulation. In the simulation, stress/strain distributions at contact locations and the skin surface deflection are quantified as cutaneous cues. In the experiments, the contact area is derived from a contiguous area on the skin surface with a super-threshold contact pressure [58], [59]. Furthermore, the deflection of the skin surface is the contour of a deflection profile in the contact plane [29], [64].

In passive touch, where compliant stimuli are indented into a fixed fingertip, a customized indenter was utilized (Fig 3.6). Participants ($n$ = 10) were instructed to rest their forearm and wrist on a stationary armrest and the index finger was constrained. Each of the four spheres (three illusion case stimuli and one distinct stimulus) was indented into the finger pad with a triangle-wave force profile peaking at the desired level (1, 2, and 3 N). To quantify the contact area at the peak magnitude of indentation, an ink-based procedure was employed. The stamped finger pad was digitized (Fig 3.6C) and the contact region was color-enhanced (Fig 3.6D). The contact areas were then calculated based on the exterior outlines with scaled pixels (Fig 3.6E).

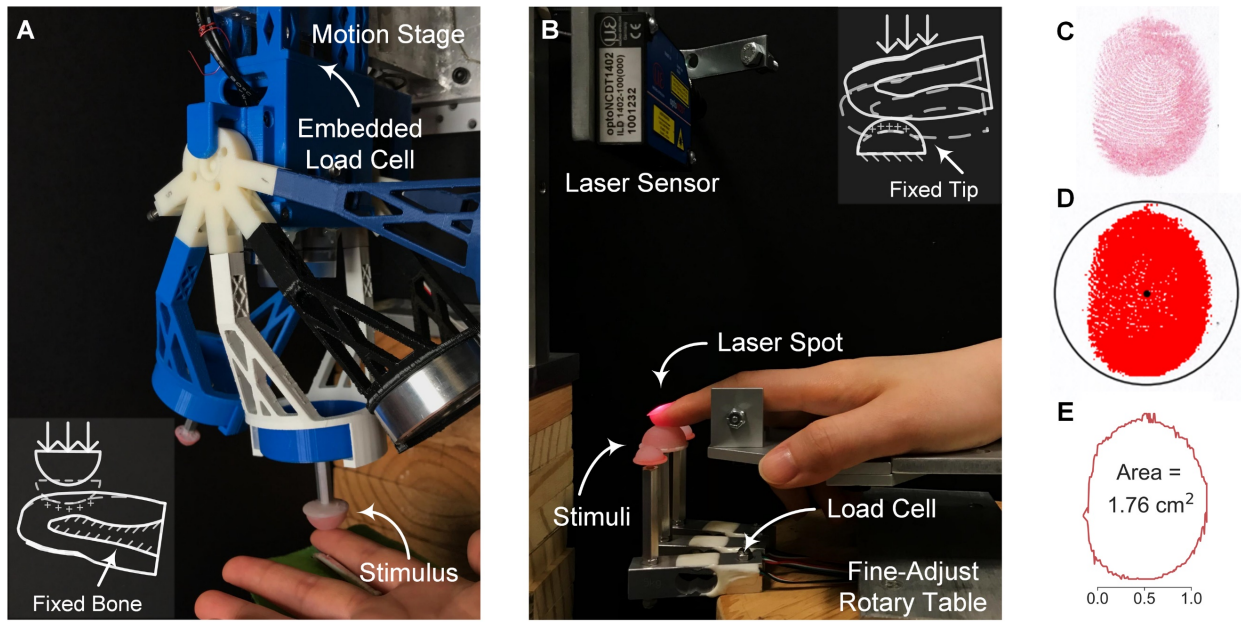

**Figure 3.6 Experimental setup and ink-based contact area analysis.** (A) For passive touch, the compliant stimulus is indented into the fixed finger pad by the motion stage. Contact force is measured by the embedded load cell. (B) For active touch, the designated stimulus is fixed and volitionally contacted by the index finger. Touch force is measured by the load cell underneath and fingertip displacement is captured by the laser sensor. (C) Contacted fingerprints are stamped and digitized for analysis. (D) The contact region is identified and color-thresholded. (E) Contact area is calculated based on the exterior outline and scaled pixels.

Non-distinct relationships of touch force and contact area are indeed observed in passive touch between the illusion case spheres across loading levels (Fig 3.7). By inspecting results from the example participant (Fig 3.7A), illusion case spheres (10 kPa-4 mm, 90 kPa-6 mm, and 90 kPa-8 mm) generated similar contact areas while the distinct sphere (10 kPa-8 mm) afforded higher contact areas. There was a significant difference between contact areas of the illusion case and distinct spheres across all force levels ($U = 0.0$, $p < 0.0001$, $d = 4.81$). In particular, data points for the three illusion cases were well clustered across all force levels (mean contact area: $0.90 \pm 0.12$ cm$^2$, mean ± SD), while the others were significantly distinct from them (mean contact area: $1.68 \pm 0.18$ cm$^2$). For all participants aggregated (Fig 3.7C), the force-contact area relation appeared to be consistent within an individual. Traces for the three illusion cases well overlapped (no significant difference detected) across all force levels, while the trace for the 10 kPa-8 mm sphere was distinct. Specifically, there was a significant difference between contact areas of the illusion case and distinct spheres across all force levels ($U = 87.0$, $p < 0.0001$, $d = 4.49$).

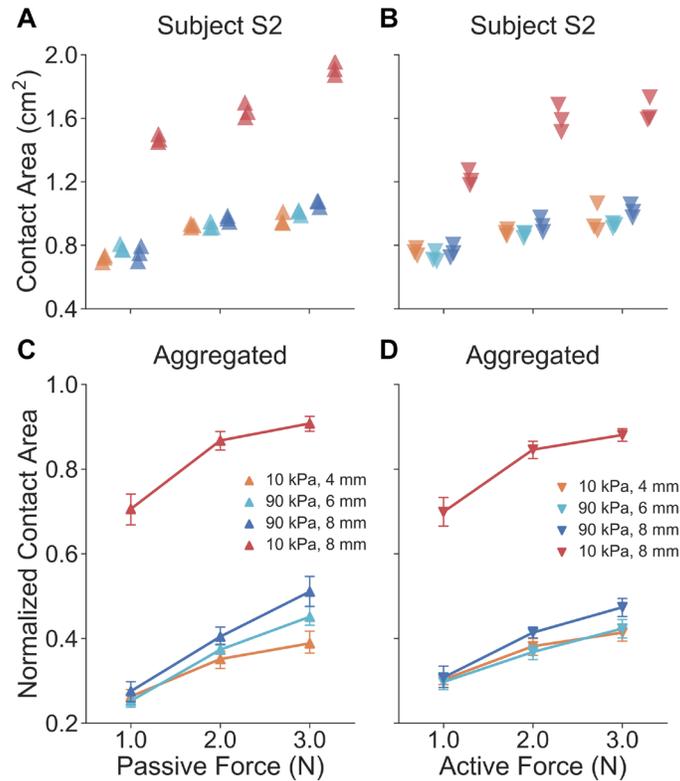

**Figure 3.7 Results of experiment 2: biomechanical measurements of contact area.** For a representative participant, in both (A) passive and (B) active touch, gross contact areas for illusion case spheres across all force levels are nearly identical, as opposed to the 10 kPa-8 mm sphere. Note that each data point represents the contact area measured from each indentation. For all participants aggregated, both in (C) passive and (D) active touch, curves of the illusion cases well overlap across all force levels, as opposed to the 10 kPa-8 mm sphere. Error bars denote 95% confidence intervals.

In active touch, where the finger volitionally touches the fixed compliant stimulus, an experimental setup was built as illustrated in Fig 3.6B. Participants ($n$ = 10) were instructed to press their index fingers down into a spherical stimulus without external constraint. A sound alarm was triggered to end each exploration when the touch force reached the desired level. After each exploration, the ink-based procedure was conducted to measure the contact area between the finger pad and stimulus.

Similar force-contact area relations were found in active touch as found in passive touch. Within a participant (Fig 3.7B), and similar to the passive touch experiments, the illusion case spheres generated similar gross contact areas while the 10 kPa-8 mm sphere exhibited higher values. There was a significant difference between results of the illusion case and distinct spheres across all force levels ($U$ = 0.0, $p$ < 0.0001, $d$ = 3.73). Specifically, the mean contact area for the three illusion cases is 0.87 ± 0.10 cm$^2$ while the other distinct stimulus derived a mean contact area of 1.48 ± 0.20 cm$^2$ across all force levels. For all participants aggregated (Fig 3.7D), traces for the illusion cases well overlapped (no significant difference detected), and the 10 kPa-8 mm sphere yields a much more distinct relationship. Specifically, there was a significant difference between the contact areas of the illusion case and distinct spheres across all force levels ($U$ = 0.0, $p$ < 0.0001, $d$ = 4.94). Since cues tied to contact area are not significantly different, proprioceptive inputs evoked in active touch may be vital to discriminating the illusion case spheres.

### 3.2.3 Experiment 3: Psychophysical Evaluation of the Elasticity-Curvature Illusion

The results of Experiment 2 support the hypothesis that cutaneous contact cues are not significantly different among illusion case spheres, for both passive and active touch. To evaluate whether there is a perceptual illusion in exploring these compliant spheres, we conducted psychophysical experiments with human-subjects.

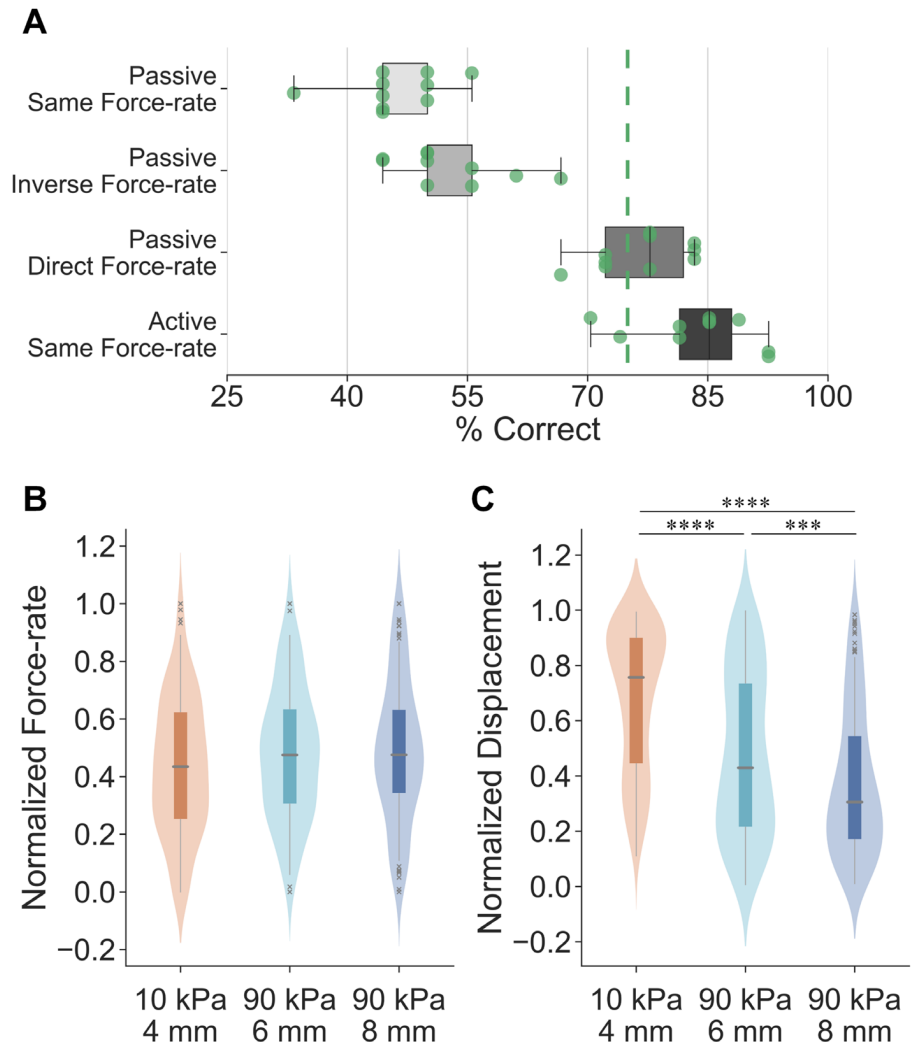

**Figure 3.8 Results of experiment 3: psychophysical evaluations and exploratory strategies.** (A) Psychophysical evaluations of illusion case spheres under different experimental conditions with all participants aggregated. The detection threshold is set as 75% for the same-different procedure. Points denote individual results. (B) Non-distinct force-rate cues are behaviorally applied for each illusion case sphere in active exploration of compliances. A miniature boxplot is set in the interior of the kernel density estimation of the underlying distribution. (C) Significantly higher fingertip displacement is applied for the small-compliant sphere, as opposed to the harder spheres. ***$p < 0.001$, ****$p < 0.0001$.

Participants ($n = 10$) were first instructed to discriminate the illusion case spheres in passive touch. To further investigate the utility of temporal cues in augmenting our discrimination performance, the indentation force-rate was systematically modulated in three different experimental conditions. In the "passive same force-rate" task, where the indentation rate was controlled at 1 N/s (Fig 3.8A), participants were not able to discriminate the stimuli (percentage of correct responses: 46.1% ± 5.7). In addition, the sensitivity measure $d'$ was also calculated under the assumption of differencing rule [69]. The mean $d'$ of 0.42 indicated a chance performance (detailed in Table 3.1). These illustrate that when only cutaneous cues are available, but their contact areas do not differ, these spheres indeed are indiscriminable.

Then, to evaluate the discriminability of these stimuli when adding proprioception to cutaneous contact, controlled force inputs were induced in passive touch in two separate cases. In the "passive inverse force-rate" task (Fig 3.8A), where the softer stimulus was indented "inversely" at a higher force-rate (2 N/s) than the harder stimulus (0.5 N/s), participants were still unable to discriminate the compliances with a percentage of correct responses of 52.8% ± 6.7. However, this result (with all participants aggregated) was significantly higher compared with the "passive same force-rate" condition ($U$ = 24.0, $p$ < 0.05, $d$ = 1.03). The mean *d'* value of 1.19 across stimulus pairs also indicated an improved, but still poor discrimination sensitivity under this condition (detailed in Table 3.1). This aligns with prior work demonstrating that participants exhibit a chance performance (~50%) when force-rate cue is "inversely" applied in passive touch [8].

**Table 3.1 Signal Detectability of the Three Illusion Case Spheres.** The sensitivity measure, $d'$, is derived from the hit and false-alarm rates, providing a bias-free measure of detectability. Under the assumption of differencing rule, $d'$ values for each condition are determined from Table A 5.4 in [69].

|  | Stimulus pair* | The signal detectability | | |
|---|---|---|---|---|
|  |  | Hit rate | False-alarm rate | Sensitivity $d'$ |
| Passive same force-rate | (10,4) & (90,6) | 0.35 | 0.33 | 0.41 |
|  | (10,4) & (90,8) | 0.43 | 0.35 | 0.84 |
|  | (90,6) & (90,8) | 0.33 | 0.38 | 0.00 |
| Passive inverse force-rate | (10,4) & (90,6) | 0.48 | 0.18 | 1.81 |
|  | (10,4) & (90,8) | 0.45 | 0.28 | 1.26 |
|  | (90,6) & (90,8) | 0.33 | 0.3 | 0.50 |
| Passive direct force-rate | (10,4) & (90,6) | 0.78 | 0.28 | 2.61 |
|  | (10,4) & (90,8) | 0.73 | 0.23 | 2.56 |
|  | (90,6) & (90,8) | 0.78 | 0.15 | 3.13 |
| Active same force-rate | (10,4) & (90,6) | 0.75 | 0.28 | 2.47 |
|  | (10,4) & (90,8) | 0.88 | 0.22 | 3.40 |
|  | (90,6) & (90,8) | 0.98 | 0.20 | 4.72 |

Third, in the "passive direct force-rate" task, where the softer stimulus was indented "directly" at a lower force-rate (0.5 N/s) than the harder stimulus (2 N/s), participants could differentiate the illusion case spheres near a 75% threshold (76.7% ± 5.4). This percentage of correct responses (with all participants aggregated) was significantly higher compared to the "passive inverse force-rate" task ($U$ = 0.5, $p$ < 0.0001, $d$ = 3.72) and the "passive same force-rate" task ($U$ = 0.0, $p$ < 0.0001, $d$ = 4.33). The values of participants' sensitivity were also improved for all stimulus pairs (detailed in Table 3.1). These results empirically validate that, when force-rate cues are "directly" applied during the contact, cues besides those cutaneous become available in discriminating the illusion case spheres. It further indicates that the controlled force-rate cues may elicit alternate perceptible inputs and are likely perceived akin to proprioception, a point which will be detailed in the Discussion.

Fourth, to validate the hypothesis that the proprioceptive cue of active finger displacement may help to discriminate the illusion case stimuli, psychophysical evaluations were conducted in active touch. Participants ($n$ = 10) were instructed to discriminate the illusion case spheres under fully active, behavioral sensorimotor control. Non-distinct force-rate cues were applied in exploring the illusion case spheres (Fig 3.8B), therefore, this experimental condition was denoted as "active same force-rate". As illustrated in Fig 3.8A, the spheres were readily discriminable with a percentage of correct responses of 83.7% ± 6.9 and a mean sensitivity of 3.53 (detailed in Table 3.1). This presents significantly better discrimination performance (with all participants aggregated) compared to the "passive direct force-rate" task ($U$ = 21.0, $p < 0.05$, $d = 1.08$). Altogether, the proprioceptive cues elicited by active, volitional control of finger movements, help in discriminating the stimuli amidst indiscriminable cutaneous contact areas.

Furthermore, in active touch, participants volitionally move their fingers to generate consistent force trajectories between stimuli (Fig 3.8B) and thereby utilize the resultant differences in the fingertip displacements between the illusion case stimuli to discriminate them (Fig 3.8C). Specifically, given the same terminal indentation force level (2 N) and non-distinct force-rate cues (no significant difference detected) among illusion cases, significantly higher displacement was applied for the softer spheres (10 kPa-4 mm vs. 90 kPa-6 mm: $U = 8786.0$, $p < 0.0001$, $d = 0.74$; 10 kPa-4 mm vs. 90 kPa-8 mm: $U = 5737.5$, $p < 0.0001$, $d = 1.18$). This finding aligns with the finite element simulation where the 10 kPa-4 mm sphere exhibited higher fingertip displacement under the same force load (Fig 3.3D). In summary, when cutaneous cues, as well as force-related movement cues, are controlled, elicited differences in fingertip displacements help discriminate the illusion case spheres.

## 3.3  Materials and Methods

### 3.3.1  Ethics Statement

The human-subjects experiments were approved by the Institutional Review Board for the Social and Behavioral Sciences at the University of Virginia. Written informed consent was obtained from all participants.

### 3.3.2  Geometry of the Fingertip Model

Two simplified 2D finite element models were derived from the geometry of a 3D model of the human distal phalanx bone [64]. The plane strain model of a cross-sectional slice from proximal first digit to distal tip was built for contact across the finger width (Fig 3.9). Meanwhile, the axisymmetric model revolving around the centerline of the finger pad was built for the contact normal to the surface (Fig 3.9). More details of the model's structure and mesh are further explained in [68].

### 3.3.3 Material Properties of the Fingertip Model

Hyperelastic material properties were used of the Neo-Hookean form of the strain energy function. The strain energy $\Psi$ was derived as:

$$\Psi = C_{10}(\bar{I_1} - 3) + \frac{1}{D_1}(J-1)^2 \qquad (3.1)$$

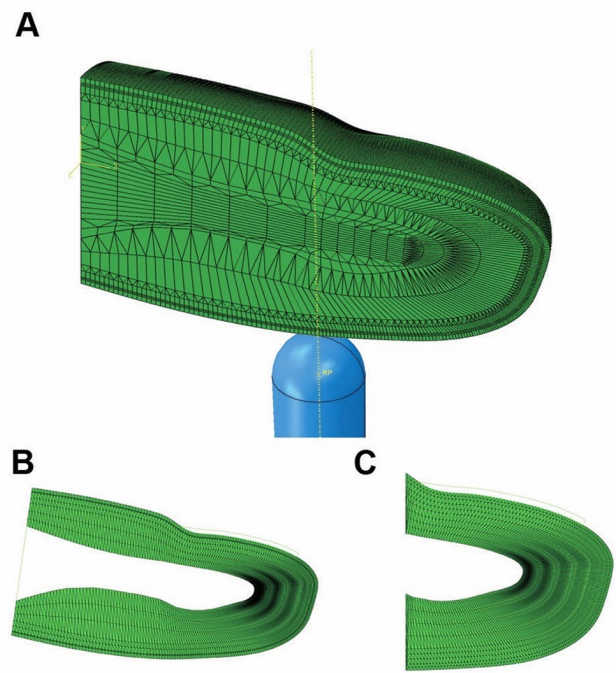

**Figure 3.9 Geometry of the finger and stimulus tip model.** (A) The compliant stimulus is implemented as hemispheres contact the skin surface of the finger pad. (B) Plane-strain model to fit the surface deflection. (C) Axisymmetric model to fit force-displacement relation and perform simulations. Adapted from [64] with permission.

where $C_{10}$, $D_1$ were material constants [64], $\bar{I}_1$ was the modified first strain invariant, and $J$ was the volume ratio known as the Jacobian matrix. The initial shear modulus $G$ was predefined and the initial bulk modulus was as $K = G/10^5$. The relationship between modulus and material constants were defined as $G = 2C_{10}$ and $K = 2/D_1$ accordingly.

The material elasticity was defined by its initial shear modulus $G$ which fully justified the material. Note that the material is in fact non-linearly hyperelastic. The Neo-Hookean model was applied to simplify the fitting procedure and derive a more robust calibration only based on the modulus $G$. Furthermore, instead of a linear Young's modulus, the hyperelastic form was considered for the soft objects which deform in a finite-strain region.

Finally, material calibration was conducted in two steps. First, the ratios of material elasticity between each layer were fitted to match the observed surface deflection to different displacements [70]. Second, the fitted ratios were scaled to fit the observed force-displacement relationships [71]. The detailed fitting procedures and final results are explained in [68].

### 3.3.4  Stimulus Tip Model

Three values of radii (4, 6, and 8 mm) and elasticity (10, 50, and 90 kPa) were selected and the stimulus tips were modeled as hemispherical with the surface of the central section attached to a rigid plate. The Poisson's ratio to the plate was set to 0.475 to mimic the nearly incompressible behavior of rubber. To suppress stress concentrations near nodes, triangular elements with 0.25 mm edge length were used in the region contacting the finger surface. Larger elements of up to 1.0 mm were used in the non-contact region to lower the computational cost.

### 3.3.5 Numerical Simulations

Nine stimulus tips (3 radii by 3 elasticity) were built based on the 2D axisymmetric model and contact mechanics were simulated in an attempt to approximate passive and active touch interactions. In passive touch (Fig 3.6A), compliant stimuli were indented into the fixed fingertip at loads of 0.25, 0.5, 1, and 2 N. The response variables were derived as cutaneous cues only, quantified by stress distributions at the epidermal-dermal interface (470 µm beneath the skin surface), calculated by averaging neighboring elements at each interface node. Note that there were in total 111 element nodes employed to cover the locations from 0 to 15.2 mm. Proprioceptive cues were decoupled since the reaction force was provided by the fixture instead of muscle activity. In active touch (Fig 3.6B), the fingertip was ramped into the fixed stimuli to the aforementioned loads. The response variables were derived from both the cutaneous and proprioceptive cues. Specifically, the proprioceptive cue was approximated by the force-displacement relation in the normal direction. This measure is tied to the change of muscle length as detected by muscle spindles, while force indicates the change of the muscle tension of Golgi tendons [41], [56], [66].

### 3.3.6 Stimuli and Experimental Apparatus

Nine compliant stimuli (3 radii by 3 elasticity) were constructed from a room temperature curing silicone elastomer (BJB Enterprises, Tustin, CA; TC-5005 A/B/C). To achieve the desired modulus, based on prior calibrations [58], corresponding ratios of cross-linker were added and mixed. These formulations were then cast into 3-D printed molds of three radii (4, 6, and 8 mm) and cured to become stimulus tips.

As illustrated in Fig 3.6A, a customized motion stage (ILS-100 MVTP, Newport, Irvine, CA) was built to indent the stimulus into the stationary finger pad [8]. Normal contact force was recorded with a load cell (22.2 N, 300 Hz, LCFD-5, Omega, Sunbury, OH) mounted onto the cantilever. The 3D printed housing fixture was equipped with a servo motor (Parallax standard servo, Rocklin, CA) and actuator arms, enabling a quick switch between different stimuli. Customized circuitry and software were developed to command the indentations. Physical measures were employed to eliminate any movement of the finger pad during the indentation. First, the participant's forearm was supported by a stationary armrest bolted onto the base of the motion stage. Velcro straps were further used to constrain the forearm if any slipperiness was detected. Second, a plastic semicircular fixture was installed to hold the index finger. The inner diameter was determined based on the dimensions of participants' distal phalanx to fasten the distal and proximal interphalangeal joints. Finally, the finger pad was held at approximately 30 degrees relative to the stimulus surface.

The experimental setup for active touch is shown in Fig 3.6B. Instrumented load cells (5 kg, 80Hz, TAL220B, HTC Sensor, China) were installed on a fine-adjust rotary table which can be rapidly rotated to present the designated stimulus. To measure the fingertip displacement, a laser triangulation displacement sensor (10 µm, 1.5kHz, optoNCDT ILD 1402-100, Micro-Epsilon, Raleigh, NC) was mounted and the laser beam was calibrated to aim at the center of the stimulus surface. The forearm, wrist, and palm base rested on a parallel beam with no external constraints.

### 3.3.7  Measurement of Contact Area

The gross contact area between the stimulus surface and finger pad was measured by the ink-based method [8], [58]. An overview of this method is shown in Fig 3.6 and summarized as follows. At the beginning of each measurement, washable ink (Craft Smart, Michaels Stores, Inc., Irving, TX) was fully applied onto the stimulus surface. After each contact, the participant was instructed to gently indent the finger pad onto a blank section of a sheet of white paper, to fully transfer the stamped ink. The remaining ink on the finger pad was then completely removed. This procedure was repeated until all measurements were completed for the participant. The sheet of paper was then marked with a 5.0 cm reference bar and digitized for analysis. A center-radius pair was selected by the analyst to identify a region enclosing the fingerprint. The desired color rendering was adjusted to outline the edges from the background. Next, a serial search was conducted to find these bounding edges and the reference bar was also identified to scale the pixels. The final area was calculated using Gauss's formula in squared centimeters.

### 3.3.8 Measurement of Force and Displacement

The gross contact readings from the force and laser sensor were smoothed to remove electrical artifacts by a moving filter with a window of 100 neighboring readings. The ramp segments of the force curves were then extracted based on first-order derivatives [9]. Linear regression was applied to the segments and the derived slope was noted as the force-rate. On the other hand, the fingertip displacement was calculated as the absolute difference between the initiation and conclusion of each movement.

### 3.3.9 Participants

The human-subjects experiments were approved by the Institutional Review Board at the University of Virginia. Ten naïve participants were recruited (5 females and 5 males, 27.5 ± 2.6 years of age) and provided written informed consent. No history of upper extremity pathology that might impact sensorimotor function was reported. All participants were right-handed and were assigned to complete both the biomechanical and psychophysical experiments. All experimental tasks were completed and no data were discarded.

### 3.3.10 Experiment Procedure

In Experiment 2, the biomechanical measurement experiments were conducted in both passive and active touch with four stimuli (illusion case: 10 kPa-4 mm, 90 kPa-6 mm, and 90 kPa-8 mm; distinct case: 10 kPa-8 mm). For passive touch, all four stimuli were each indented into the finger pad at three force levels (1, 2, and 3 N) respectively. Each stimulus was ramped into the finger pad for one second and retracted away for one second. The ink-based procedure was applied for each indentation. There were three indentations for each stimulus at each indentation level per participant. All indentations were separated by a 20-second break. For active touch, the four stimuli were palpated by the index finger at three force levels which were behaviorally controlled. In particular, participants were instructed to actively press into the designated stimulus and a sound alarm was triggered to end the current exploration when their force reached the desired level. The ink-based procedure was used for each exploration. There were three explorations for each stimulus at each force level per participant. All explorations were separated by a 20-second break.

In Experiment 3, psychophysical discrimination experiments were conducted for both passive and active touch with the three illusion case stimuli. Following the rule of ordered sampling with replacement, nine stimulus pairs were drawn from the three illusion case spheres and were prepared for psychophysical evaluation. The stimulus ordering within each pair was determined by the sampling results (see Table 3.1 for detailed assignments). Participants were blindfolded to eliminate any visual information about the stimulus compliance or the movements of the indenter and the finger pad. No feedback on their performance was provided during the experiment. Using the same-different procedure, after exploring each pair (one touch per stimulus), participants were instructed to report whether the compliances of the two were the same or different. Note that the same-different procedure was applied herein because the observer can use whatever cues are available and does not have to articulate how the compliances actually differ [19], [72]. This fits with the scope where the roles of perceptual cues are under investigation.

For passive touch, each trial consisted of discriminating one stimulus pair. Following the sampling order, spheres from the same pair were ramped into the fixed finger pad successively (Fig 3.6A). The indentation interval was controlled as 2-seconds to obtain consistent temporal effects on perception [21]. All discrimination trials were separated by a 15-second break. The terminal force level was set to 2 N as this aligned with Experiments 1 and 2. As illustrated in Fig 3.8, three experimental tasks were performed in passive touch. In the "passive same force-rate" task, all stimuli were indented at 1 N/s to 2 N. In the "passive inverse force-rate" task, a higher force-rate was applied for the soft stimulus while the lower force-rate was applied for the hard stimulus. The 10 kPa-4 mm sphere was indented at 2 N/s to 2 N. The 90 kPa-6 mm and 90 kPa-8 mm spheres were indented at 0.5 N/s to 2 N. In the "passive direct force-rate" task, force-rate was applied in a direct positive relation with the stimulus modulus. The 10 kPa-4 mm sphere was indented at 0.5 N/s to 2 N and the two 90 kPa spheres were indented at 2 N/s to 2 N. For each experimental task, each of the nine stimulus pairs was presented twice. Adapted from prior studies [19], [21], the test order of all trials was randomized to balance the carry-over effects in response bias [73].

For active touch, the experiments were conducted under participants' fully active, behavioral control (Fig 3.6B). Within each discrimination trial, a participant was instructed to explore compliance by palpating each of two spheres successively with a terminal touch force of 2 N. When their force reached 2 N, a sound alarm was triggered to end that exploration. The interval between two explorations was set to 2-seconds as previously noted. Force and fingertip displacement were recorded simultaneously. Each stimulus pair was presented three times in a randomized order to balance the carry-over effects in sequential responses. There was a 15-seconds break between trials. Note that trials under the same experimental task were grouped together and conducted within one block. The test order of the four experimental tasks (blocks) was randomized for each participant.

### 3.3.11 Data Analysis

As illustrated in Figs 3.7 and 3.8, the experimental results for all participants were aggregated for analysis. A normalization procedure was required for data aggregation since participants exhibited distinct sensorimotor capabilities, range of finger movements, and dimensions of the finger pad [8], [9]. In particular, for each experimental task, all recordings of each tactile cue were normalized to the range of (0, 1) by sigmoidal membership function [6], [9]. The center of the transition area was set as the mean value of the data normalized, and the logistic growth rate of the curve was set to 1. After this transition was completed for each participant, all results were then aggregated together for statistical analysis. The Mann–Whitney U test ($\alpha$ = 0.05, two-sided test) was applied to compare the samples and the Cohen's *d* (the absolute value) was calculated for statistically significant results to evaluate the effect size. The confidence interval was derived by bootstrapping the estimated data with 1000 iterations.

## 3.4 Discussion

This study investigates an illusion phenomenon in exploring soft objects, specifically the situation in which small-compliant and large-stiff spheres are indiscriminable. These two physical attributes are common to everyday tasks; for example, in judging the ripeness of fruit. Through a combination of solid mechanics modeling, biomechanical contact measurement, and psychophysical evaluation, we show that small-compliant and large-stiff spheres afford nearly identical cutaneous contact, and thus, are indiscriminable in passive touch where only cutaneous cues are available. However, this phenomenon vanishes in active touch, when proprioceptive cues augment indiscriminable cutaneous contact cues. Furthermore, the results indicate that in exploring compliant objects, force-controlled movements are more efficient and optimal for eliciting the cutaneous and proprioceptive cues that underlie our judgments of compliance.

**A force-control movement strategy is optimal, efficient, and underlies softness perception.** Amidst indiscriminable cutaneous contact cues, participants behaviorally control the exploratory forces they apply to soft objects. Specifically, the terminal indentation force, as well as the rate change of touch force was behaviorally controlled to be non-distinct among the illusion case spheres (Fig 3.8B). Indeed, participants actively move their fingers to apply consistent force trajectories and thereby evoke significant differences in fingertip displacement cues for softness discrimination. These fingertip displacements are proprioceptive by nature and critical to the discrimination of the illusion case stimuli (Fig 3.8C). Indeed, this exploratory strategy is important from a number of other perspectives. First, a force-modulation strategy is essential to compensate for the natural remodeling of the skin over time, which leads to changes in its thickness and elasticity [29]. Such changes in the skin's mechanics could generate a large variance in neural firing patterns, and thereby perception. However, the skin can reliably convey information about indentation magnitude, rate, and spatial geometry when touch interactions are controlled by surface pressure. Since force directly converts to pressure on the skin upon contact, a force-modulation strategy echoes theories of active, behavioral control when exploring soft objects in daily tasks [23], [29]. Second, at the behavioral level, we prioritize exploratory force to optimize our perception of object compliances in relevant contexts [23], [65], [74], [75]. Indeed, the availability of force-related cues improves discriminability by reducing the necessary deformation of the skin [8]. Similarly, for the exploratory procedure of pinch grasp, we control the grip force within a safety margin, informed by skin mechanoreceptors, to prevent slipping or applying exceedingly high pressure [76], [77].

**Change of cutaneous contact as a cue to proprioception.** As just discussed, the force-control movement strategy is efficient and optimal in evoking differentiable cues, in active touch. In passive touch, we observe that participants can discriminate the illusion case stimuli, particularly in the "passive direct force-rate" case, with a percentage of correct responses of about 77% (Fig 3.8A). While lower than the discrimination result for active touch, this represents a significant improvement over the "passive same force-rate" case, which yields chance performance.

We hypothesize that the modulation of force under "passive direct/inverse force-rate" condition – where the softer stimulus was indented "directly/inversely" at a lower/higher force-rate than the harder stimulus – provides an alternate perceptible input during the dynamic contact phase, also tied to finger proprioception. In particular, in alignment with prior findings [8], [9], we show that the rate of change of force is linearly correlated with the rate of change of gross contact area [68]. While we cannot directly measure the rate of change of contact area, due to the limitations of the ink-based method only being able to measure terminal contact area, one could easily extrapolate this correlation to the dynamic contact phase by discretizing the terminal contact area/force into the instantaneous contact area/force. Using the 3D imaging technique, we indeed demonstrated that force-rate cues can proportionally elicit the instantaneous change of contact area [14]. Such cues might therefore induce the illusion of fingertip displacement amidst dynamic contact [66], [78]. In particular, Moscatelli, *et al.* demonstrated that skin deformation of this kind naturally induces a sensation of relative finger displacement in the stationary hand [56], [79]. Similarly, stretching the skin at the proximal interphalangeal joint can induce illusions of self-motion in anesthetized fingers [78]. Moreover, microscopic oscillatory stimulation at the skin surface also can elicit illusory finger displacements when pressing on a stiff surface [80]. Therefore, when passively exploring the illusion case spheres under the modulation of force-rate, the improved discriminability is likely derived from the proprioceptive sensation elicited by the change of contact area, which is originally induced by the force-rate cue.

**A perceptual illusion phenomenon inspired by everyday tasks.** The stimulus attributes of elasticity and curvature can be found in everyday, ecologically relevant tasks, e.g., judging the ripeness of fruit for edibility. In some prior studies, however, stimuli have been highly engineered and delivered by sophisticated devices [46]. Such stimuli may not afford the same perceptual acuity as ecologically accurate soft objects [43]. Moreover, stimulus compliance at times has been parameterized by its stiffness rather than its modulus [5], [7], [64], which can be confounding for naturalistic objects of identical stiffness but differing in geometry [64]. Herein, we address these issues by building spherical stimuli with covaried radii and elasticity which recapitulate important properties of ecologically compliant materials and mimic the contact profile of the skin surface's contacting elastic objects [45], [46], [58]. As it is difficult to measure the material properties of the fruit, which can break down rapidly between sessions, our group has begun to consider the perceptual commonality between silicone-elastomer materials as reasonable stand-ins for ecological fruits [9]. Similar to the work with engineered substrates herein, we have found that the exploratory strategy of behaviorally controlling force aligns with how we judge the ripeness of fruit. In particular, we volitionally pinch soft fruit, by controlling grip force, to help differentiate their ripeness [9].

**Computational modeling formulates psychophysical studies.** Instead of evaluating empirically with human-subjects a large number of stimulus combinations of elasticity and curvature, we computationally identified combinations with indistinct cutaneous contact. Indeed, a "computation first" effort as such demonstrates an alternative paradigm to bridge theoretical and empirical studies, make specific predictions, and test particular hypotheses. Specifically, to better understand the encoding mechanism underlying the identified tactile illusion, cutaneous and proprioceptive cues need to be dissociated. As this is empirically demanding, we employed two interaction modes (passive and active touch, Fig 3.6) in the computational simulation. The potential cues and interaction modes that modulate the illusion are then validated in psychophysical experiments with human-subjects.

Finally and relatedly, far fewer illusions have been discussed in the tactile modality than for vision and audition [45], [46]. This partially reflects the fact that tactile illusions are not as easily accessible [45]. Indeed, sophisticated efforts are usually required to create appropriate conditions to conceive the illusion, which is a significant electromechanical challenge to achieve empirically [46]. The "computation first" approach demonstrated herein may help in identifying potential illusions more efficiently.

## 3.5 Acknowledgments

We would like to thank all the participants of the human-subjects experiments, as well as members of the Gerling Touch Lab for fruitful discussions and feedback. The content is solely the responsibility of the authors and does not necessarily represent the official views of the National Institutes of Health or the National Science Foundation.

# 4 Aim II. Decipher Optimal Strategies and Perceptual Modeling in Tactile Explorations among Populations

## 4.1 Introduction

In our activities of daily living, we frequently interact with soft, compliant, and deformable objects. For example, we might judge the ripeness of a fruit or touch the arm of a friend to offer comfort [81], [82]. In more specialized environments, such as surgery, physicians may seek to distinguish tissue and ducts from fat and bone or palpate abnormalities. Such daily interactions require us to judge, recognize, and discriminate among individual objects and groups of objects. We do so by perceptual procedures which integrate and update our prior expectations, exploratory strategies, elicited sensorimotor inputs, and internal representations [6], [44], [83].

Many ongoing efforts are refining our understanding of the physical and perceptual cues that help encode our sense of compliance. One paradigm is that we rely upon cues of cutaneous contact as a function of touch force. Distinct efforts have focused upon cutaneous cues of contact area, skin deformation, and spatial distribution of pressure [5], [7], [8], [57]. However, recent studies have shown that the terminal contact area is not readily discriminable, and additional cues likely augment our judgments of compliance [6], [14], [58], [64]. Among those, our proprioceptive system provides vital inputs through the kinesthetic sense of joint angles [6], spatiotemporal patterns in cutaneous contact [56], and visual-haptic integration [61], [84], [85]. For instance. The rotation about the metacarpophalangeal (MCP) joint is considered to reliably encode finger penetration into compliant objects and help fine-tune our sensorimotor control of movements [85], [86]. That said, exactly how we modulate our exploratory motions to obtain optimal perceptual cues remains unclear. We do know that people tend to apply higher forces as a robust strategy to obtain higher differential sensitivity. They also tend to utilize steeper finger angles against the stimulus surface when discriminating harder objects [23], [24]. Overall, these physical cues are recruited and integrated into multimodal signals, fine-tuned by optimal exploratory strategies, which modulate sensorimotor movements [23], and then transferred to perceptual space where compliances may be discriminated. Strategies likely vary between person, task, and compliance range. That said, certain exploratory movements are more optimal than others and can more efficiently link with and elicit certain perceptual cues, especially those mediated by interactions with naturalistic objects where our tactile system has been finely tuned.

However, most current studies of compliance interactions use engineered materials, such as silicone-elastomers, foams, and robotic devices to stand-in for ecological materials. This is done because of the difficulty of acquiring, controlling, and quantifying the properties of naturally occurring soft objects such as animal or plant tissue, both from sample to sample and over time. To this point, in naturalistic settings, there have been very few studies of human interaction with ecological stimuli. For example, Katz studied bakers in their occupational interactions with dough and like substances, breaking down dimensions such as stickiness and elasticity [87]. Weber, et al. employed surface textures associated with fabrics, e.g., velvet, fleece, and drapery tape [88]. Another effort assessed the visual and tactile perception of the shape of bell peppers [89]. Recently, Cavdan, *et al.* considered perceptual dimensions of a very wide array of soft naturalistic objects and their associated exploratory procedures [43]. In a different field of study, robotics researchers have sought to sense differences in the properties of fruits and other naturalistic objects to sort, grasp, and manipulate them [90], [91].

In summary, we do not yet understand which exploratory strategies and perceptual rules elicit which perceptual cues that most optimally encode the material compliance of both engineered and ecologically relevant objects. As a step in this direction, this work uses a multi-measurement approach in studying contact interactions, exploration strategies, and perceptual modeling with: a) soft plum fruit in both single finger touch and two finger grasp; b) bare finger palpation with compliant spheres where stimulus elasticity is decoupled from the radius of curvature.

## 4.2 Materials and Methods

The work herein studies human perception alongside physical contact interaction cues in the discrimination of both naturally-occurring compliant objects, the plum, and engineered compliant spheres. Psychophysical experiments were performed alongside biomechanical measurements, where participants operated under their own active, volitional control of their finger movements. The resultant physical contact and interaction cues were analyzed.

For the study with the soft fruit, the plum was chosen because it is commonplace, can be grasped in one hand, and represents a naturalistic task in which compliance discrimination might determine relative ripeness. Methods were developed and adapted for use with this natural object. In particular, an ink-based method was used to capture the finger pad to stimulus contact area, and force sensors and a non-contact laser captured imposed force and fingertip displacement. Setups were built such that plum interactions could be measured for single finger touch and thumb-index pinch grasp. In a human-subjects study with thirteen participants, five pairs of plums were used, each with one riper than the other. Several considerations in the experiment's design were made to accommodate the readily perishable and easily damaged fruit specimens.

For the study with the compliant spheres, experimental setup and methods were adopted and modified from Aim I. In particular, active exploration of object softness was conducted with bare finger touch, based on the experimental platform shown in Fig 3.6B. Biomechanical measurement was performed alongside psychophysical evaluation, where touch force and finger movement were captured by multimodal sensors. In discriminating softness of stimuli pairs, the exploration duration required for one trial was systematically constrained and controlled from minimal to infinite, such that changes in perceptual strategies adopted by participants could be captured, modeled, and validated.

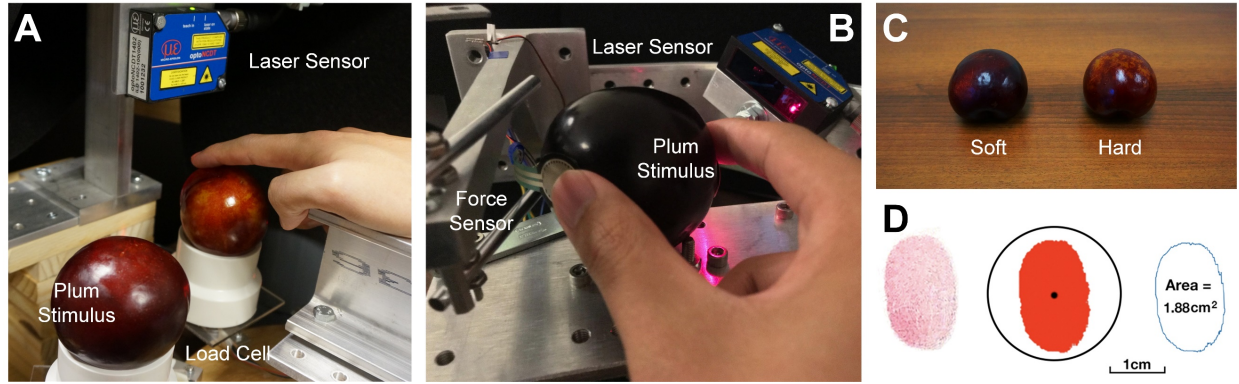

**Figure 4.1 Experimental setup of test rigs and ink-based contact area analysis.** (A) In the setup for single finger touch, a set of two platforms is installed on a rotary table to present the designated plum. Imposed force is measured by a load cell underneath each platform. The position of the fingertip is monitored by a laser displacement sensor. (B) The setup for thumb-index pinch where the plum is grasped and held in a horizontal orientation. Grasp force is the average reading of two force-sensitive resistors which are firmly attached to the plum's surface and are contacted by the finger pad(s). The position of the index fingertip is tracked by the laser sensor. (C) One pair of soft and hard plums. (D) Plum contacted fingerprints are stamped onto a sheet of white paper and then digitized. The contact region is identified and color-enhanced. Contact area is calculated using Gauss's formula based on the exterior outline and scaled pixels.

### 4.2.1 Experimental Apparatus for Exploring Soft Fruits

For the grounding condition of single finger touch, an experimental test rig was built as shown in Fig 4.1A. Specifically, two platforms were installed on a fine-adjust rotary table that can be rapidly rotated into position to present the designated stimuli. Each of the platforms was fitted with a circular pipe upon which the spherical object was set and could be rotated. An instrument load cell (5 kg, HTC Sensor TAL220B, China) was mounted beneath each platform to measure the imposed force at 80 Hz. A non-contact laser displacement sensor (optoNCDT ILD 1402-100, Micro-Epsilon, Raleigh, NC) was mounted above to measure the position of the participant's fingertip nail at 1.5 kHz. The participant's forearm and wrist rest on a beam parallel to the table surface without constraints.

For the condition of thumb-index pinch grasp, a test rig was built as shown in Fig 4.1B. The plum was held by four carriage bolts which were fixed on the platform so that participants could grasp the plum on the bolts. Force-sensitive resistors (0.1-10 kg, Interlink Electronics 400/402, Camarillo, CA) were firmly attached on symmetrical sides of the plum and contacted by the thumb and index finger pads in measuring touch force. The position of the index finger was measured by the laser sensor, mounted horizontally.

### 4.2.2 Naturalistic Soft Stimuli

Ten plums were employed in the experiments. Of several varieties available, the pluot (Honey Punch Pluot) was selected, which is a hybrid fruit developed in the late 1980s that is 75% plum and 25% apricot. They resemble plums with smooth skin and similar shapes and originate from California, USA. The perceptual effects brought by surface roughness and texture can be eliminated due to the smooth surface of the plum. The plums were about 6.3 ± 0.2 cm in diameter. A brief psychophysical experiment was conducted to establish the basic discriminability of the plums. This experiment used an insertion sort procedure to evaluate the compliances, resulting in a sorted plum array from the hardest $P_1$ to the softest $P_{10}$. The first five were denoted as the "hard" plums, with the last five as the "soft" plums. Finally, five pairs consisting of one hard and one soft plum were assigned accordingly as: ($P_1$, $P_6$), ($P_2$, $P_7$), ($P_3$, $P_8$), ($P_4$, $P_9$), and ($P_5$, $P_{10}$). In the subsequent experiments, each pair was used by one of the five participant groups respectively, and those experiments were completed within 24 hours of purchase. The number of times a location on the plum successively touched was controlled to avoid irrevocable damage to the plums.

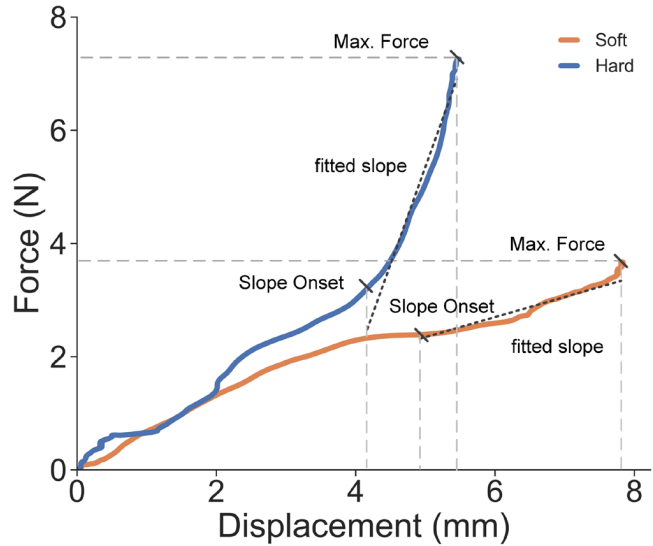

**Figure 4.2 Relationships of touch force and fingertip displacement for exemplar soft and hard plums for one trial.** Linear regression is applied to the segments from slope onset to the peak to obtain the rate of change of the curve, defined as the observation of fitted virtual stiffness.

### 4.2.3 Measurement of Displacement and Force Rate

As illustrated in Fig 4.1A, the emitted laser was aimed at the surface of the participant's fingernail to monitor its position. Readings from the laser sensor were smoothed to remove any electrical artifacts by a moving average filter with a window of 100 neighbor values. Fingertip displacement was derived by the absolute difference between initiation and conclusion of the finger movement. On the other hand, to calculate force and force-rate, readings from the force sensor were first filtered by the aforementioned filter. The ramp segments were then extracted according to the first-order derivatives. As reported previously, the ramp onset and ending were defined based on the peak derivatives [6]. Finally, a linear regression was applied to the ramp and the slope was noted as the force-rate.

### 4.2.4 Measurement of Virtual Stiffness

Stiffness describes the resistance of an elastic object to deformation or deflection by an applied force. In active exploration of compliant objects, physically perceived stiffness – i.e., the relationship between force and displacement – could be utilized to encode the perception of compliance [83]. The overall estimate of virtual stiffness $K$ was derived by a data fusion procedure where two individual observations of the perceived stiffness were combined through the Kalman filter: $K = x_1 + \sqrt{\sigma_1^2/(\sigma_1^2 + \sigma_2^2)}(x_2 - x_1)$ where $x_{1,2}$ and $\sigma_{1,2}$ denoted the mean and standard deviation of aggregated experimental results of these two observations respectively. The first observation from the peak point was derived as the maximum applied force divided by the corresponding fingertip displacement, as done with a simple physical model in tapping. The second observation was derived from the fitted slope. As shown in Fig 4.2, the ramp of the force-displacement curve was extracted by the same procedure as aforementioned. Linear regression was applied to obtain the rate of change of the ramp, defined as another observation of virtual stiffness.

### 4.2.5 Measurement of Contact Area

An ink-based method developed in prior work was applied to measure the gross contact area between the plum surface and finger pad at various levels of imposed force [8], [92]. An overview of the basic steps is shown in Fig 4.1D and detailed as follows. Washable ink (Craft Smart, Michaels Stores, Inc., Irving, TX) was fully applied onto the plum surface at the beginning of each trial to ensure there is always a sufficient amount left on the surface. After contact with the plum, the participant was instructed to gently indent the finger pad onto a sheet of plain white paper, fully transfer the stamped ink. Before each new trial, the remaining ink was completely removed from the finger pad. After all trials were completed, each individual's sheet of fingerprints was digitized and processed with a 5.0 cm scale line. Custom software was used for analysis. In particular, the analyst identifies a center point within each fingerprint, as well as the region enclosing the fingerprint. The desired color threshold is selected to enhance the image. Next, a serial search is conducted within the identified region of interest to outline the area with boundary pixel points. After determining the length of each pixel in centimeters via the scale line, the final area is calculated using Gauss's formula and scaled in squared centimeters.

### 4.2.6 Data Normalization

To aggregate experimental results across all participants, a data normalization procedure is required because of distinct perceptual capabilities among participants. For each task, all recordings of a measured tactile cue from each participant were first normalized to the range of (0,1) by the sigmoidal membership function [6] and then aggregated across all participants. The value of the sigmoid midpoint was set as the average of the data normalized, and the sigmoid function's slope was set to 1.

### 4.2.7 Participants

The study was approved by the Institutional Review Board at the University of Virginia. In total, 13 healthy subjects were recruited to participate in this study (8 females, 5 males, mean age = 24.8, SD = 2.0). According to the Edinburgh-handedness inventory, all participants were right-handed [93]. All participants provided informed consent and were assigned to five plum pairings consisting of 3, 2, 2, 2, and 4 members, respectively. The first four groups completed Task 1-4 and the last group completed the second part of Task 4 only. The third participant in group 1 completed only Task 1-3 because the experimenters were concerned that the plum had been irrevocably damaged and would not yield reliable results. Additionally, note that the data on fingertip displacement and virtual stiffness from group 1 are not shown because the finger (due to its small size) moved out of the range of the laser on some trials. This went unnoticed until after all tasks were completed and 4.9% of trials were discarded.

### 4.2.8  Experimental Procedure

**Task 1 – Biomechanical measurement with behaviorally controlled force in single finger touch.** To measure the biomechanical relationship between imposed force and contact area in single finger touch, this task was designed where force levels (2, 4, and 6 N) were behaviorally controlled and presented during three sessions respectively. The test order of the sessions was randomized to balance the effects of fatigue or inattention. Participants were instructed to press the index finger down into the stimulus and a sound alarm was triggered to end the trial when the force reached a predefined constraint. The contact area was then measured by the ink-based method. For each participant in group 1, there were two trials for each plum at each force level, for a total of 36 trials. For the other groups, there were three trials for each plum at each force level, for a total of 108 trials. All trials were separated by 20-second.

**Task 2 – Psychophysical discrimination with fully active touch in single finger condition.** Psychophysical discrimination of three combinations of the two plums in one pair (soft-hard, soft-soft, and hard-hard) was conducted under the participants' fully active, volitional control. Combinations were presented in a randomized order. Participants were blindfolded to remove visual information about the plum ripeness and their finger movements. Using the same-different procedure, after exploring both stimuli from one combination, participants were instructed to report whether the compliances of the two were the same or different. Force and displacement were recorded simultaneously. Each participant completed three trials for each combination, for a total of 81 trials. All trials were separated by a 30-second break.

**Task 3 – Psychophysical discrimination with fully active touch in pinch grasp.** The same three combinations for each plum pair were randomly presented in this task. Participants were instructed to use the thumb and the index finger to pinch the plum on the fixed bolts horizontally, as illustrated in Fig 4.1B. After exploring both plums, participants reported whether the compliances of the two plums were the same or different. Each participant completed three trials for each combination, for a total of 81 trials. All trials were separated by a 30-second break.

**Task 4 – Biomechanical measurement with behaviorally controlled force in pinch grasp.** In the first part, the biomechanical relationships between force and contact area were measured at three behaviorally controlled force levels (low, medium, and high) which were presented during three randomly ordered sessions. Participants pinched the presented plum at their volitional control and notified the experimenter to end the current trial when the grasp force reached the desired level. The contact area was then measured by the ink-based method. For group 1, there were two trials for each plum at each force level, for a total of 24 trials with two participants. For the other groups, there were three trials for each plum at each force level per participant, for a total of 108 trials. All trials were separated by a 20-second break.

In the second part, the biomechanical measurement of grasp force and index finger displacement were conducted at two force levels (low and high). Force sensors were contacted and covered by the finger pads over the course of one trial. Participants notified the experimenter to end the current trial when they perceived the applied grasp force has reached the desired levels. There were 10 trials for each plum at each force level, for a total of 80 trials with two participants. For the other two participants, there were 9 trials for each plum at each force level, for a total of 72 trials. All trials were separated by a 20-second break.

### 4.2.9 Modeling Perceptual Strategy of the Differencing Rule

In the first perceptual strategy, to have a robust estimation of the minimum time for discriminating the compliance between sequentially explored plums, the accumulations of force were extracted and processed for similarity analysis. As shown in Fig 4.3A, selected trial data were averaged and cropped for the soft and hard plums, respectively, resulting in two force curves over the same duration. Corresponding force-rate curves were also calculated by the aforementioned method. The discriminability between the soft and hard curves is quantified by the discrete variation of the Fréchet distance which is a measure of the similarity between polygonal curves [94]. To improve computational efficiency, each force curve was downsampled by a factor of 50, resulting in two polygonal curves $H:[0,n]$, and $S:[0,n]$ while each was made of *n* connected segments, as illustrated in Fig 4.3C. The sequences of the endpoints of the line segments were denoted by $\sigma(H) = (u_1,\ldots,u_n)$ and $\sigma(S) = (v_1,\ldots,v_n)$ respectively. A coupling *L* between *H* and *S* was defined as a non-decreasing sequence of distinct pairs from $\sigma(H) \times \sigma(S)$ as follows:

$$(u_{a_1}, v_{b_1}), (u_{a_2}, v_{b_2}), \ldots, (u_{a_m}, v_{b_m}) \tag{4.1}$$

where $a_1 = b_1 = 1, a_m = b_m = n,$ and for each $i \in \{1,2,\ldots,m-2\}$, let $a_{i+1} = (1-\lambda)a_i + \lambda(a_i + 1)$, $b_{i+1} = (1-\lambda)b_i + \lambda(b_i + 1)$ with all $\lambda \in \{0,1\}$. The length $\|L\|$ was then calculated as the maximum Euclidian distance among those sequence pairs:

$$\|L\| = \max_{j=1,\ldots,m} d(u_{a_j}, v_{a_j}). \tag{4.2}$$

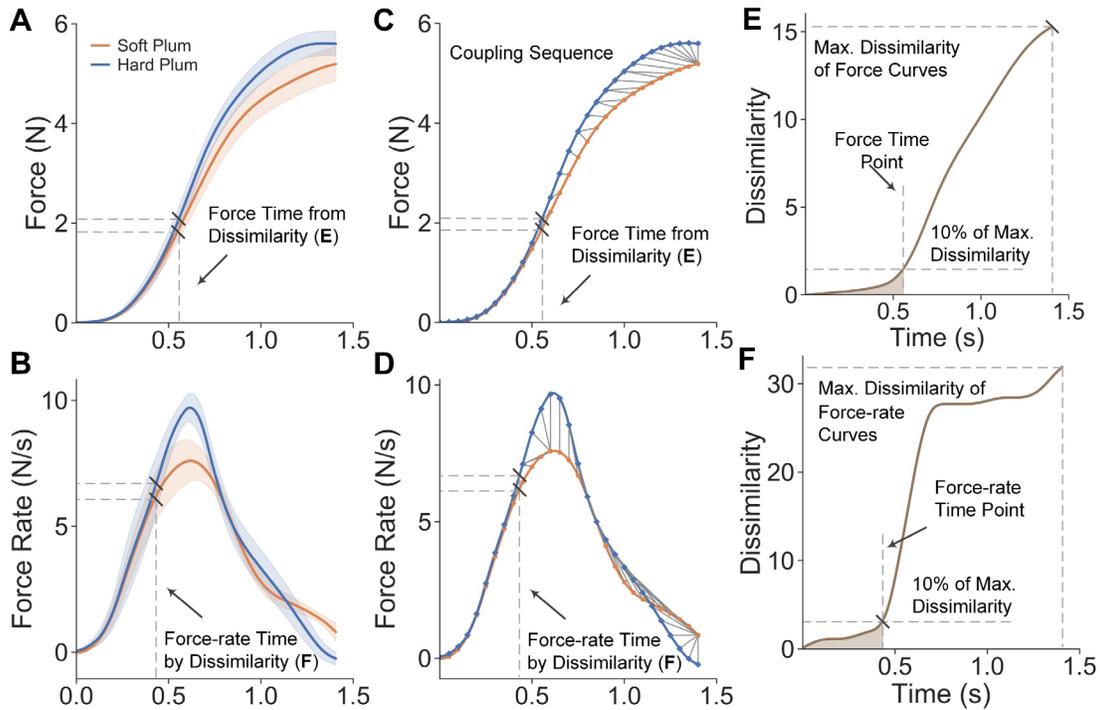

**Figure 4.3** Modeling the perceptual strategy for discriminating compliances between sequentially explored plums by differencing force-related cues with exemplar trial data. (A) Touch force curves over a partial course of exploration for the soft-hard plums. Translucent bands denote the standard deviation of force values. (B) Corresponding force-rate curves for soft and hand plums. Translucent bands denote the standard deviation of force-rate values. The coupling sequences are illustrated along with each discrete distance between endpoints of segments for the force curves (C) and corresponding force-rate curves (D). The similarity between the soft-hard curve is quantified by the discrete Fréchet distance for the force (E) and force-rate cue (F). The detection threshold is set according to the reported JNDs of force perception and the final estimates of the required minimum time for discriminating the soft-hard plums are denoted accordingly.

Finally, the discreet Fréchet distance between curve *H* and *S* was calculated as

$$\delta_{dF}(H,S) = \min\{\|L\|\} \tag{4.3}$$

which grows positively from zero as the two curves become more dissimilar, as illustrated in Fig 4.3E. Based on the just noticeable differences (JNDs) for human perception of touch force reported by prior studies [65], [95], a threshold of 10% was set to find the time point when one could differentiate force cues from the representations of the soft and hard plums, which was defined herein as the estimate of required minimum time for compliance discrimination. As shown in Fig 4.3D and 4.3F, the time estimate with the force-rate cue was also calculated by the aforementioned method accordingly. Therefore, two different estimates of the required minimum time over the course of compliance discrimination were derived by differencing force and force-rate cues.

## 4.2.10 Modeling Perceptual Strategy of the Independent-Observation Rule

In the second perceptual strategy, to have a reliable estimation of the minimum time for independent recognition of the compliance retained and gathered within single exploration, force-displacement curves were extracted and processed by the proposed model based on the Kalman filtering procedure. As illustrated in Fig 4.4, selected trial data were averaged and cropped for the soft and hard plums respectively, resulting in two curves with the same length $n$. Considering a noise-free linear spring model in tapping, the instantaneous virtual stiffness (excluding for the first time point) was derived as

$$k_j = \tan \alpha_j = F_j / d_j \tag{4.4}$$

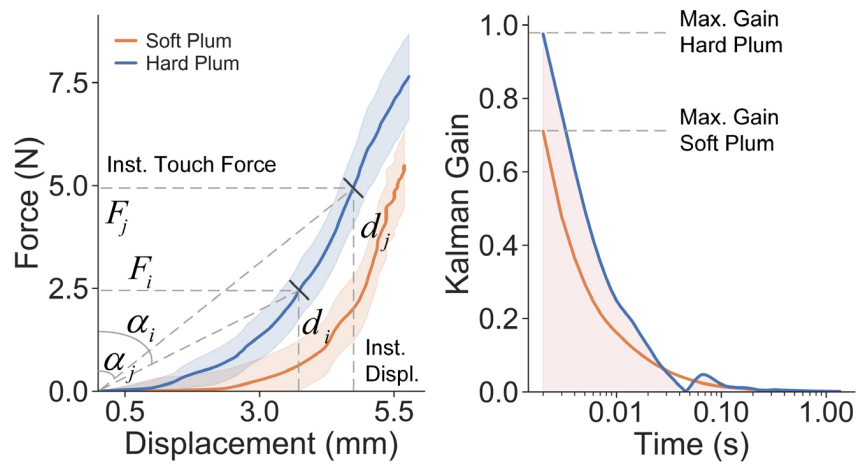

**Figure 4.4** Modeling the perceptual strategy for the independent-observation of compliance based on the updating model of virtual stiffness cue with exemplar trial data. Left: Relationships of touch force and fingertip displacement over a partial course of exploration. Instantaneous virtual stiffness is computed as the tangent of angle α. Translucent bands denote the standard deviation. Right: The Kalman gain quantifies the robustness of compliance estimates over the course of model updating. The x-axis is set to be logarithmical for more details. Detection thresholds are set as 10% of the maximum gain to find the required minimum time for reliable recognition of each plum compliance over the duration of exploration.

according to Hooke's law for each $j \in \{2,3,\ldots,n\}$. Adapted from the recursive Bayesian updating model proposed in [83], the estimate of the virtual stiffness $\hat{K}_i$ was initiated by $k_2$ and updated with stepwise inputs from virtual stiffness cues. Each step $i \in \{3,4,\ldots,n\}$ updated prior estimate $\hat{K}_{i-1}$ by combining weighted input of current virtual stiffness $k_i$:

$$\hat{K}_i = \hat{K}_{i-1} + K_g(k_i - \hat{K}_{i-1}). \tag{4.5}$$

$K_g \in (0,1)$ denoted the Kalman gain which was derived by the covariance of prior and current estimate:

$$K_g = \sqrt{\sigma_{i-1}^2 / (\sigma_{i-1}^2 + \sigma_i^2)}. \tag{4.6}$$

Finally, the stiffness estimate evolved recursively over time with an updated covariance $\sqrt{(1-K_g)\sigma_{i-1}^2}$. As shown in Fig 4.4, the Kalman gain approaches to zero when the estimate is updated to be stable over the course of exploration. Based on the aforementioned justification, a detection threshold of 10% was set to find the first time point when one could have a reliable percept of stiffness, which was defined herein as the estimate of the required minimum time for the recognition of the plum compliance by recursively integrating perceptual gains from virtual stiffness cues within a single exploration.

## 4.3 Results

### 4.3.1 Biomechanical Cues in Single Finger Touch

The biomechanical relationship of touch force, contact area, and fingertip displacement in single finger touch were measured at three force levels. As shown in Fig 4.5, within each participant, gross contact areas for the soft and hard plums were overlapped to be non-distinct across all force levels. Differences between the two participants were mostly due to the individual dimensions of their finger pads. With all participants aggregated, gross contact areas were overlapped to be non-differentiable, as shown in Fig 4.6. This indicated that participants could not rely only upon gross contact area cues in differentiating the compliances of the soft and hard plums.

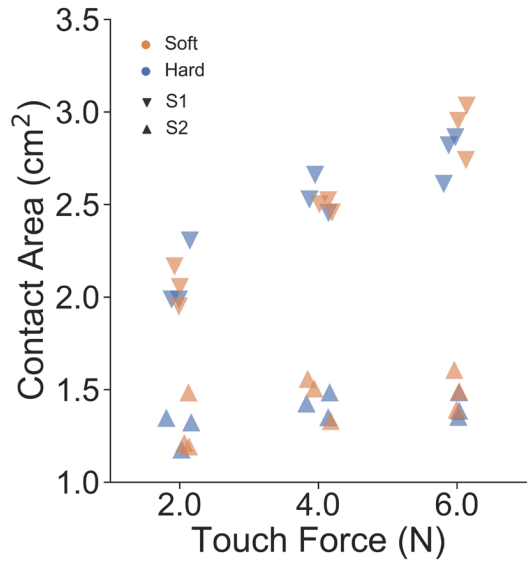

**Figure 4.5** For the index finger touch, biomechanical relationships of touch force and gross contact area for the soft-hard plum from two participants.

As shown in Fig 4.6, fingertip displacements were well separated for the soft and hard plums. Participants applied significantly higher fingertip displacements for the soft plums as opposed to the hard ones (2 N: $t_{(17)}$ = -6.929, $p < 0.0001$; 4 N: $t_{(17)}$ = -3.210, $p < 0.01$; 6 N: $t_{(17)}$ = -2.244, $p < 0.01$). This indicated that, in the behaviorally-controlled condition, when touch force is volitionally controlled to be the same, participants could still control their finger movements to elicit significantly different displacement cues between soft and hard plum pairs. These results are in line with prior work done with engineered stimuli [6], [58], [64].

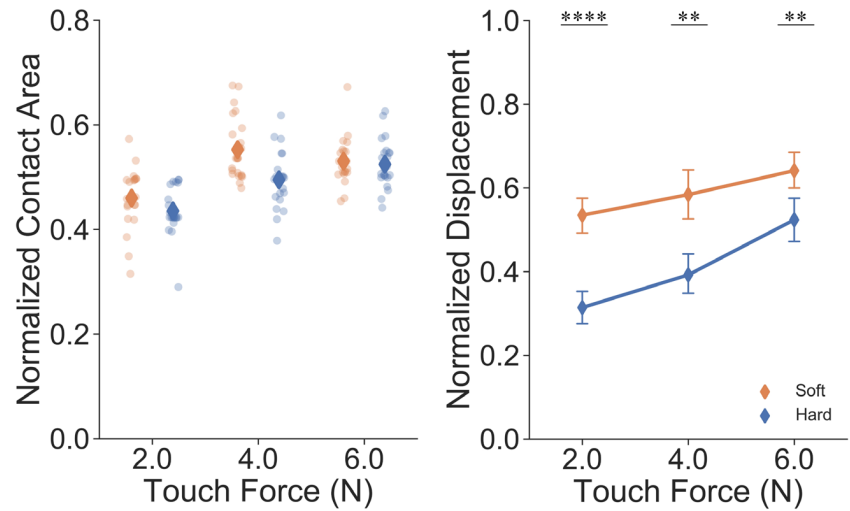

**Figure 4.6 For the single finger touch, biomechanical relationships of force, contact area, and displacement for the soft and hard plums.** Left: Normalized contact area and force with all trials aggregated. Points denote the trial data and diamonds denote the means. Right: Normalized displacement and force with all trials aggregated except for the first group. The **significance and ****significance are denoted at p < 0.01 and p < 0.0001 by a paired-sample t-test. The Cohen's *d* values are -1.79, -1.18, and -0.81 respectively. Error bars denote 95% confidence intervals.

### 4.3.2 Perceptual Cues in Psychophysical Discrimination with Single Finger Touch

Perceptual cues of force and displacement were measured for single finger touch. As shown in Fig 4.7, participants applied significantly higher peak force ($t_{(80)}$ = 5.725, $p < 0.0001$) and force-rate ($t_{(80)}$ = 2.871, $p < 0.001$) for the hard plums. Note that the aggregated $R^2$ value for the force-rate fit was 0.98 ± 0.01 (mean ± SD). In contrast, similar fingertip displacements were applied in discrimination. This indicated that in fully active exploration, participants tended to volitionally control their movements to obtain similar displacement and applied discriminable force-related cues between the soft and hard plums. This finding aligns with prior work showing the same strategy in discriminating the compliances of man-made stimuli [6], [23].

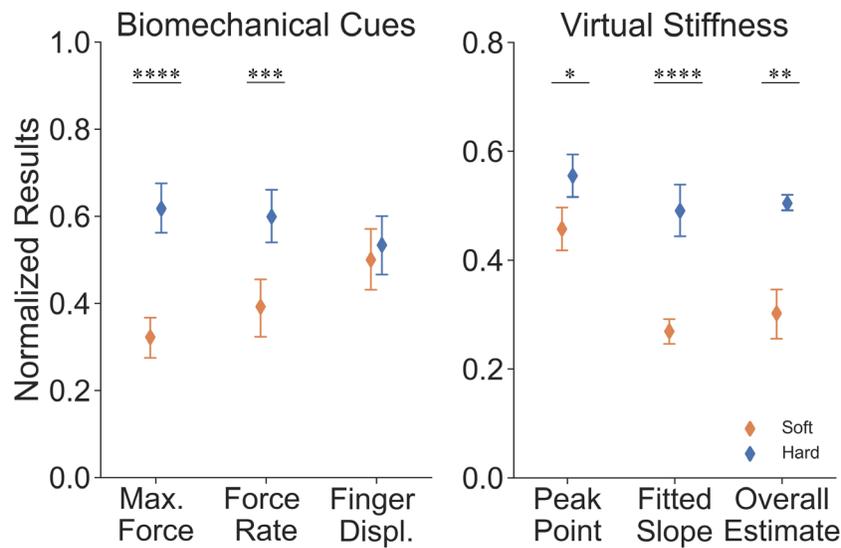

**Figure 4.7 For the single finger touch, perceptual cues in discriminating the soft-hard plums with all participants except for group 1**. Left: Normalized peak force, force-rate, and displacement. Right: Normalized observations of peak point, fitted slope, and an overall estimate of the virtual stiffness. The *significance, **significance, ***significance, and ****significance are denoted at $p < 0.05$, $p < 0.01$, $p < 0.001$, and $p < 0.0001$ by a paired-sample t-test. The Cohen's *d* values of the significant results are 1.15, 0.67, 0.45, 1.04, and 2.94 respectively. Error bars denote 95% confidence intervals.

As shown in Fig 4.7, virtual stiffness was calculated for the soft and hard plums. For the observations of the peak point ($t_{(80)}$ = 2.289, $p < 0.05$) and fitted slope ($t_{(80)}$ = 4.695, $p < 0.0001$), significantly higher stiffness values were obtained for the hard plums. Note that the aggregated $R^2$ value for the fitted slope was 0.89 ± 0.09. The overall estimate yielded the same result by the data fusion procedure ($t_{(5)}$ = 5.628, $p < 0.01$). These results indicated that the human perceived stiffness, as quantified in our measurement setup via virtual stiffness, indeed aligns with the actual compliance of the stimuli. Furthermore, the metric of virtual stiffness gives insight into the exploratory strategies, in particular, the relationship between applied force and fingertip displacement in discriminating compliances. Participants indeed volitionally control their fingertip displacement to obtain differentiable quantities of force.

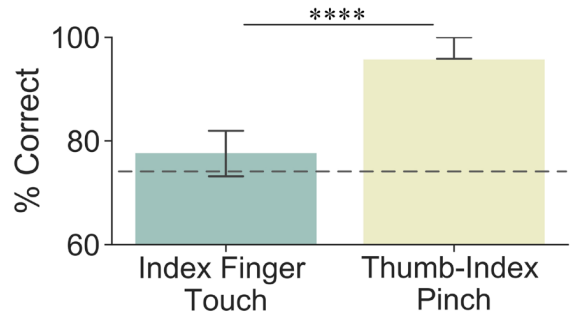

**Figure 4.8 Psychophysical discrimination in the soft-hard plums with all participants aggregated.** The discrimination threshold is set as 75 %. The ****significance is denoted at p < 0.0001 by a paired-sample t-test. Cohen's *d* value is -1.11. Error bars denote 95% confidence intervals.

### 4.3.3 Psychophysical Discrimination in Two Conditions

As shown in Fig 4.8, when discriminating with index finger touch, participants were able to differentiate the soft and hard plum with a threshold detection rate of 77.8%. Under pinch grasp, participants were able to improve their performances significantly with a correct response rate of 95.8% ($t_{(8)}$ = -3.290, *p* < 0.0001). This indicated that compared to the single finger touch, with additional perceptual cues evoked by the pinch grasp, participants could achieve better discrimination performance more naturally.

### 4.3.4 Biomechanical Cues in Thumb-Index Pinch

As shown in Fig 4.9, biomechanical relationships of force and contact area were measured at three force levels. Compared to the single finger touch, with the stabilization support from the thumb, a greater terminal contact area was obtained. However, for both fingers, the contact areas for soft and hard plums were still overlapped to be non-distinct. This indicated that the terminal contact area is not vital to discrimination, independent of force. The improved performance in pinch grasp may likely result from other perceptual inputs.

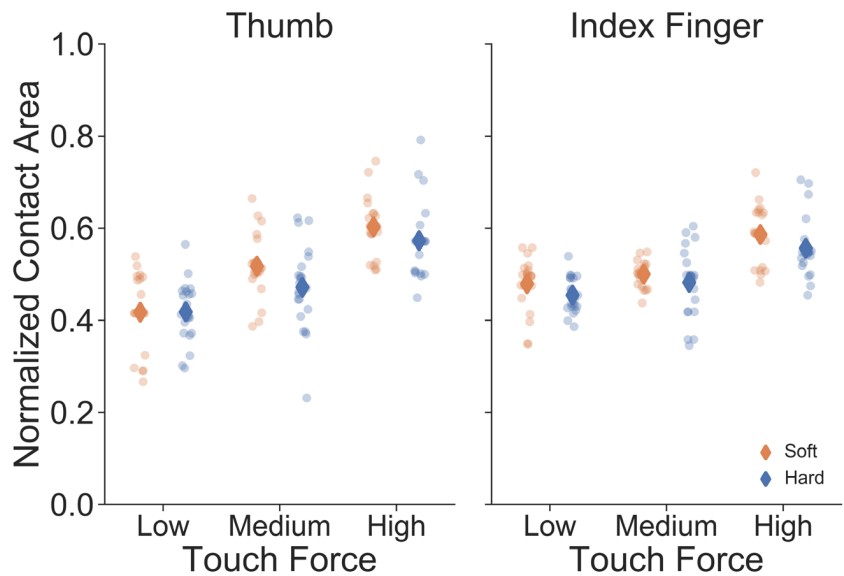

**Figure 4.9 For the thumb-index pinch grasp, biomechanical relationships of touch force and contact area in the thumb (left) and index finger (right) for the soft and hard plums with all trials aggregated.** Points denote the data from each trial and thin diamonds denote the mean values.

As shown in Fig 4.10A, a significantly higher peak force was applied to the hard stimuli across all force levels (Low: $t_{(37)}$ = 6.751, $p < 0.0001$; High: $t_{(37)}$ = 17.554, $p < 0.0001$). For the low force level, peak force was significantly lower compared to the high level (Soft: $t_{(37)}$ = 11.096, $p < 0.0001$, Hard: $t_{(37)}$ = 13.062, $p < 0.0001$). This indicated that participants could indeed behaviorally control their movements to accurately impose touch forces according to the instructions. This result in part aligns with prior work demonstrating that the indentation force is related to the experimenter's instructions during active haptic exploration [24]. As shown in Fig 4.10B, significantly higher displacement was applied for the soft plum (Low: $t_{(37)}$ = -5.375, $p < 0.0001$; High: $t_{(37)}$ = -14.575, $p < 0.0001$). As shown in Fig 4.10C, significantly higher virtual stiffness was obtained for the hard plums (Low: $t_{(37)}$ = 6.850, $p < 0.0001$; High: $t_{(37)}$ = 8.524, $p < 0.0001$). Note that the aggregated $R^2$ value for the virtual stiffness fit was 0.94 ± 0.12. However, there is no significant difference for virtual stiffness across different force levels (Soft: $t_{(37)}$ = -3.790, $p = 0.395$, Hard:

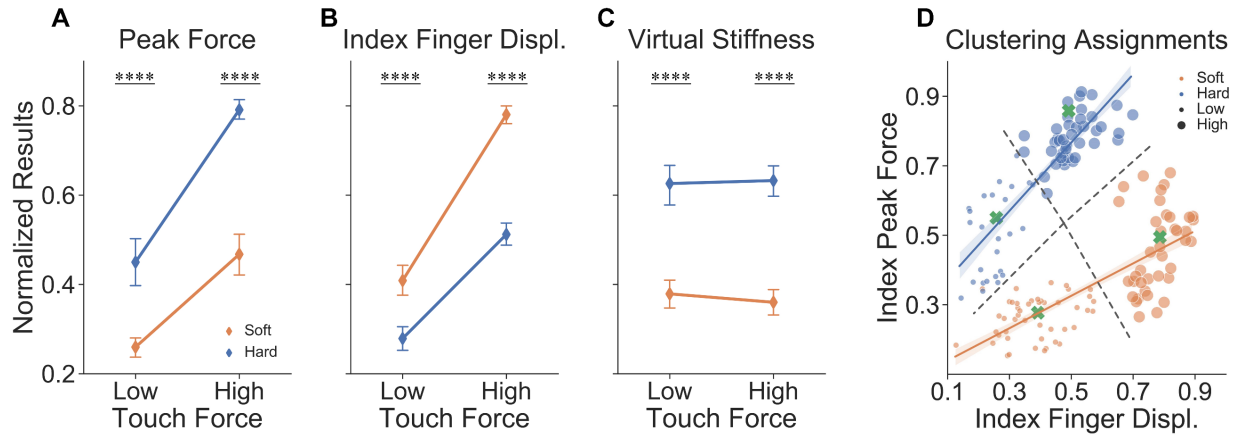

**Figure 4.10 For the thumb-index pinch, biomechanical measurements of physical cues and multidimensional clustering analysis.** Normalized results of peak grasp force (A), fingertip displacement (B), and overall estimate on virtual stiffness (C) for the soft and hard plums with all trials aggregated. The ****significance is denoted at p < 0.0001 by a paired-sample t-test. Error bars denote 95% confidence intervals. The Cohen's *d* values of the significant results are 1.67, 3.52, -1.33, -3.84, 2.01, and 2.69 respectively. (D) All trial data are partitioned into four exclusive clusters based on peak grasp force and displacement. Centroids and dashed lines indicate the cluster regions. Linear regression procedures are applied to visualize the correlation on clustered data. Translucent bands denote 95% confidence intervals for regression estimations.

$t_{(37)}$ = 0.285, *p* = 0.820). This indicated that virtual stiffness is a reliable measure to quantify the compliance of the plums, independent of touch force. Indeed, we observe, in the psychophysical discrimination with a single finger, that participants volitionally control fingertip displacement to perceptually differentiate force.

Multidimensional clustering analysis was conducted to verify which perceptual cue could optimally discriminate the compliances. As shown in Fig 4.10D, plum compliances could not be differentiated solely by peak grasp force or fingertip displacement. In contrast, the combination of these two cues could partition all trial data into four exclusive groups by the *k*-Means algorithm. The match rate between original and clustered data was 92.1%. Each group represented a combination of plum compliance (soft or hard) and peak force (low or high), which was partitioned into the four cluster regions. Linear regression procedures were applied to the clustered data. The Spearman's rank coefficient yields correlations of 0.86 ($p$ = 1.01e-20) and 0.77 ($p$ = 6.24e-18) for the hard and soft plums respectively. These statistics quantify particular correlations between peak force and fingertip displacement cues by which different compliances may be encoded. Together, they reinforce that the virtual stiffness cue could afford discrimination by solely correlating force and displacement.

### 4.3.5 Exploration Time Estimates of the Differencing Rule

Estimates of the minimum time required for compliance discrimination by differentiating force-related cues were calculated for two exploratory procedures. As shown in Fig 4.11, compared with the pinch grasp, participants tended to require significantly more time to discriminate with single finger touch when differentiating the force cues. The same trend was also obtained when employing the force-rate cue. It indicated that a more natural gesture could facilitate haptic exploration by reducing the required minimum time for differencing the force-related cues among sequentially explored stimuli.

Multidimensional clustering analysis was conducted to verify whether the gesture (exploratory procedure) indeed impacts the minimum time for discrimination via force-related cues. As shown in Fig 4.11, based on time estimates from force and force-rate, data points were correctly clustered into two groups by the $k$-Means algorithm. When moving from single finger touch to pinch grasp procedure, the correlation between time estimates of force and force-rate has been changed as indicated by the linear regression results. This indicated that time estimates of discrimination could well encode the impact brought by different exploratory procedures, and pinch grasp indeed required lesser time for discrimination by force cues.

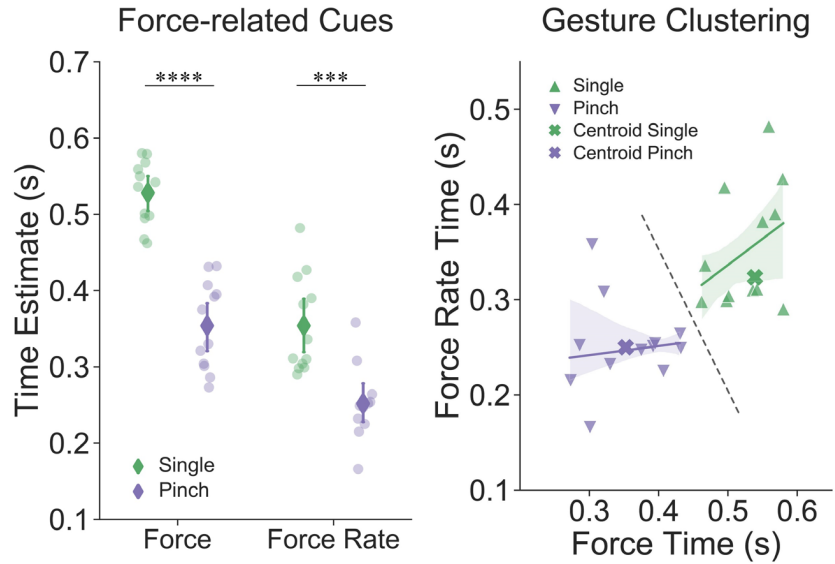

**Figure 4.11 Time estimates of sequential compliance discrimination by differentiating force-related cues and multidimensional clustering analysis on gestures with all participants aggregated.** Left: Time estimates by imposed force and force-rate cues. Points denote the results from grouped trials and diamonds denote the means. Error bars denote 95% confidence intervals. The ***significance and ****significance are denoted at $p < 0.001$ and $p < 0.0001$ by the Mann-Whitney U test. Right: All the time estimates based on imposed force and force-rate cues are partitioned into two exclusive clusters representing different gestures. Centroids and the dashed line indicate the arrangement of clusters. Linear regression is applied to illustrate the correlation on clustered data. Translucent bands denote 90% confidence intervals.

### 4.3.6 Exploration Time Estimates of the Independent-Observation Rule

Estimates of the minimum time required for compliance recognition by the integration of perceived stiffness were calculated by virtual stiffness cues for the two exploratory procedures. As shown in Fig 4.12, compared with exploring hard plums using the single finger touch, participants tended to require significantly more time to recognize the compliance of soft plums. The same result was obtained when exploring with the pinch grasp procedure. This indicated that higher compliance could facilitate active exploration by reducing the required time for compliance recognition.

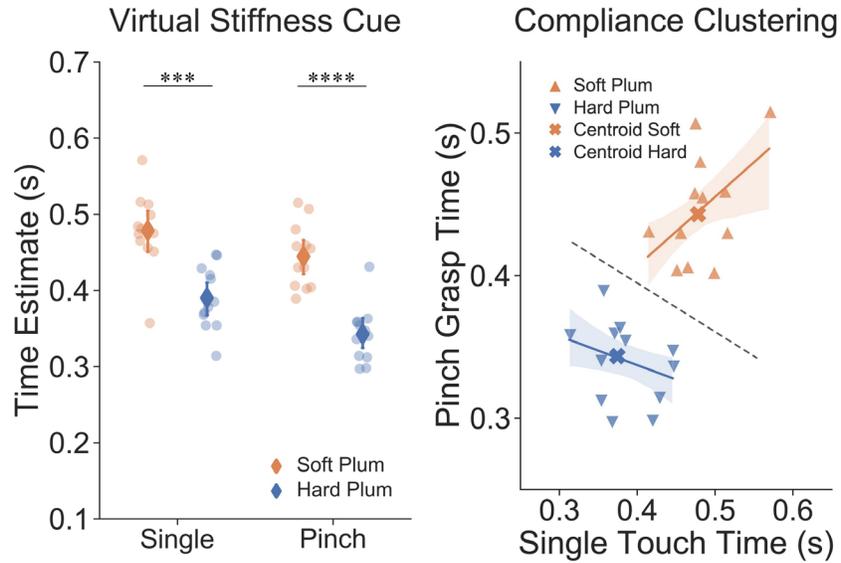

**Figure 4.12 Time estimations of perceptual integration by virtual stiffness cue and multidimensional clustering analysis on compliance with all participants aggregated.** Left: Time estimates by virtual stiffness cue for single touch and pinch grasp condition. Points denote the results from grouped trials and diamonds denote the means. Error bars denote 95% confidence intervals. The ***significance and ****significance are denoted at $p < 0.001$ and $p < 0.0001$ by the Mann-Whitney U test. Right: All the time estimates for single touch and pinch grasp are partitioned into two exclusive clusters representing different softness of the plum stimuli. Centroids and the dashed line indicate the arrangement of clusters. Linear regression is applied to illustrate the correlation on clustered data. Translucent bands denote 90% confidence intervals.

Multidimensional clustering analysis was conducted to verify whether the plum compliance indeed impacts the required minimum time for compliance recognition by the updating procedure of virtual stiffness cues. As shown in Fig. 5, based on time estimates of single finger touch and pinch grasp, data points were clustered into two groups with a matching rate of 91.7%. When exploring soft plums, the correlation between the two estimates was changed compared to the result of hard plums. This indicated that time estimates of recognition could well encode plum compliances, and hard plums indeed required lesser time for compliance recognition.

### 4.3.7 Modeling the Utility of Differencing Rule amidst Varied Exploration Time

To estimate the utility of the differencing rule employed in discriminating softness pairs, traces of touch force and force-rate cues were compared by the similarity analysis. The derived curvature dissimilarity indices for each stimulus pair were then used to indicate the absolute differences between two sensory observations and corresponding percepts.

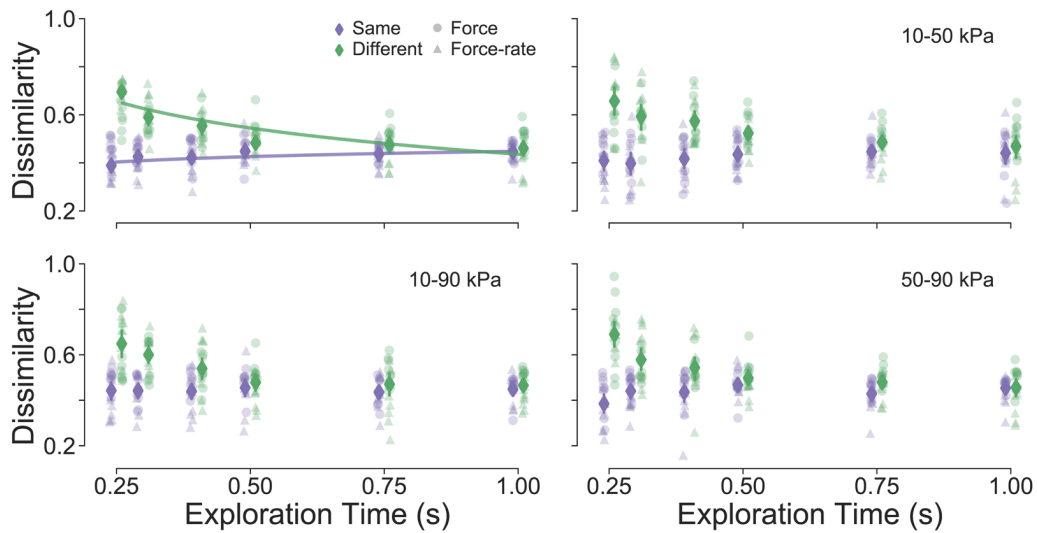

**Figure 4.13 Similarity analysis on force-related cues for the perceptual strategy of differencing rule.** For each stimulus pair category (10-50, 10-90, and 50-90 kPa), dissimilarity indices were derived both for pairs with the same and different stimuli, force and force-rate cues. Points denote the results from all participants and diamonds denote the means. Aggregated means across stimulus pairs were fitted with the logarithmic function.

As shown in Fig 4.13, with extremely limited exploration time (~0.4 s), distinct dissimilarity on tactile cues was obtained in discriminating soft stimulus pairs. In particular, similar force cues (low dissimilarity) were derived in discriminating the same stimuli, and sufficient differences in force were evoked for different stimuli in one pair. This indicates that, within limited exploration time, the strategy of the differencing rule could be employed by eliciting distinct force cues, so that the absolute sensory differences between two stimuli could afford optimal discrimination. In contrast, with sufficient exploration time (~0.75 s), similar force cues were evoked in discriminating all stimulus pairs, as the dissimilarity overlaps to be lower for both the same and different stimuli. This indicates that, when exploration time is sufficient for discrimination, the differencing rule seems to be not feasible as the absolute sensory differences are lower. Overall, when the exploration time is controlled from limited to adequate, the utility of differencing rule may be adapted from viable to be impractical.

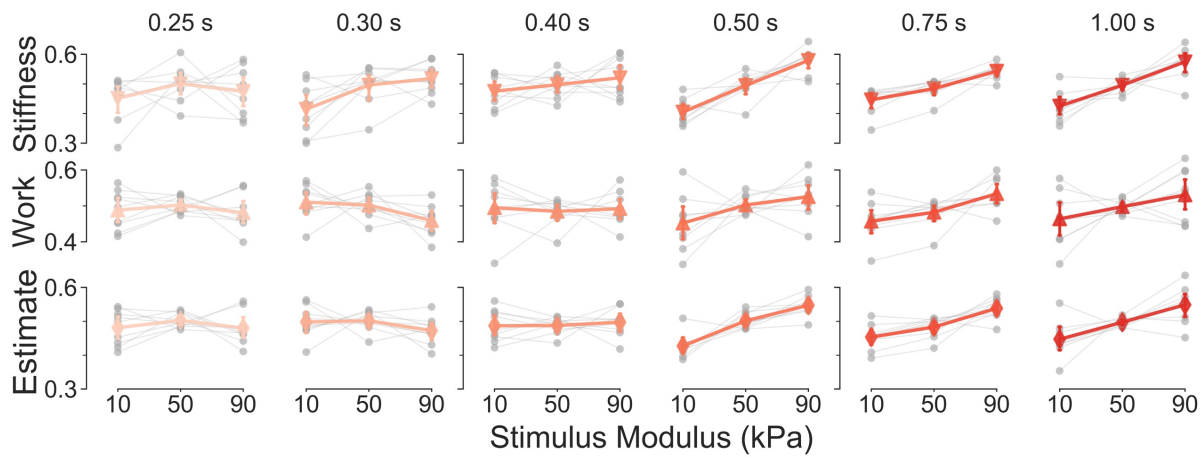

**Figure 4.14 Categorical analysis on recognition cues for the perceptual strategy of independent-observation rule.** For each stimulus sphere (10, 50, and 90 kPa), tactile cues of virtual stiffness, applied work, and integration of these two were derived from discrimination. Shaded curves denote the results from individuals and diamonds denote the means.

### 4.3.8 Modeling the Utility of Independent-Observation Rule amidst Varied Exploration Time

To estimate the utility of the independent-observation rule employed in discriminating softness pairs, tactile cues of virtual stiffness, applied work, and integration of these two were derived to correlate with explored softness. Relationships between stimulus moduli and magnitude of cues indicate how these sensory cues may encode the independent recognition on each explored stimulus.

As shown in Fig 4.14, with extremely limited exploration time (~0.4 s), those evoked recognition cues cannot accurately encode compliances among the three explored stimuli. In particular, magnitudes of tactile cues were not positively proportional to the stimulus moduli. This indicates that, within limited exploration time, the strategy of the independent-observation rule may not aid in discriminating, considering that the elicited recognition cues cannot fully encode explored compliances in the perceptual space. In contrast, with sufficient exploration time (> 0.5 s), recognition cues were elicited in a way such that the magnitudes aligned well with the explored softness. Strong positive correlations were derived between stimulus modulus and sensory recognition. This indicates that, when exploration time is sufficient for discrimination, the independent-observation rule may afford optimal discrimination since elicited recognition cues well encode stimulus compliance among explorations. Overall, when the exploration time is controlled from limited to adequate, the utility of the independent-observation rule could be adapted from impractical to be feasible.

## 4.4  Discussion

The work herein – to the authors' knowledge – is the first of its kind to quantify touch interaction cues and exploratory strategies that drive our perception of naturalistic and ecological interactions, in particular, the palpation of soft plum fruit. Nearly all prior work to study the human perception of compliance, upon direct bare finger contact, has been performed with silicone-elastomers and foams. To enable such study, sophisticated measurement techniques and experimental designs were developed and adapted for work with these natural objects. Overall, we find that gross contact area cues were non-differentiable for the soft-hard plums. In contrast, touch force and fingertip displacement differed significantly. These two variables were coupled into the virtual stiffness cue, which quantified the perceived compliances and differed significantly in discrimination (Fig 4.7). The newly defined metric of virtual stiffness illustrates how volitional strategies of exploratory movement may be tuned to generate discriminable perceptual cues. In fully active exploration, participants tended to move their fingers to particular displacements to elicit differences in reaction force, driving the discrimination of naturalistic compliances (Fig 4.7). We also noted that discrimination improved significantly for the more natural interaction of pinch grasp, despite non-differentiable gross contact areas (Fig 4.8). Indeed, for judging the differences in ripeness between the fruit stimuli, in addition to the gross contact area, virtual stiffness cues likely augment discrimination.

Noteworthy, there are no prior studies that evaluate the equivalence of perceptual strategies with silicone-elastomers, foams, and other engineered stimuli to ecologically naturalistic materials. Interestingly enough, however, we do find that engineered stimuli are, in general, reasonable approximations to these ecologically compliant objects. In particular, there was a consensus on the role of peak force as a reliable perceptual cue [6], [8], [23], [24]. Moreover, exploratory strategies described in prior studies aligned with the exploration of the soft fruit. Especially, when discriminating soft-hard plums, participants tended to volitionally match their displacement cues so as to elicit discriminable force-related cues (Fig 4.7), although displacement cues could also be utilized when force cues are behaviorally controlled to be non-distinct (Fig 4.6). Similar findings were reported when terminal contact area cues are non-differentiable [6]. Such exploratory strategies can be quantified by the virtual stiffness which maps the estimate of perceived stiffness from the physical relation between touch force and fingertip displacement. As shown in Fig 4.10C, virtual stiffness can accurately quantify the perceived stiffness and map it to the actual compliance of the plums. Further clustering analysis reinforced that the integration of force and displacement – quantified by the virtual stiffness cue – could optimally afford discrimination between plum fruit, independent of the touch force one imposes.

Several aspects regarding the experimental design could be taken into account for future work due to inherent difficulties in working with delicate objects that change over time, such as ripe fruit. Finally, the plum represents one instantiation of a natural object of our daily interactions. Others as well are of interest, including tissues of the body, amongst others. More effort is needed to consider these and associated cues and ties back to the perception of compliance, and how to represent such natural objects with robust representations akin to silicone-elastomers and foams.

Our time estimates (~370 ms) are about ten times higher in absolute magnitude compared to time estimates of first neural spikes elicited in discriminating stimuli [10], [96]. However, given neuromuscular time constraints (~160 ms) [10], multimodal integration delay (~100 ms) [62], and limitations of memory retrieval (up to 30 s) [21], our estimations at the behavioral level are reasonably sound. Moreover, our findings are derived from aggregated results between participants and sequential explorations. To gain further clarity on the utility of the two strategies, further trial-by-trial analysis is required. Finally, temporal cues of skin deformation were not utilized due to the measurement limitations in the grasp of the force-sensing resistors.

## 4.5 Acknowledgment

This work was supported in part by grants from the National Science Foundation (IIS-1908115) and the National Institutes of Health (NINDS R01NS105241). The content is solely the responsibility of the authors and does not necessarily represent the official views of the NSF or NIH.

# 5 Aim III. Individual Differences in Skin Mechanics, Exploratory Strategies, and Perceptual Sensitivity

## 5.1 Introduction

We develop individual differences in touch acuity and exploratory procedures from daily interactions in our natural environment [30], [97]. For instance, individuals respond distinctly when discerning delicate textures using the fingertips [17], employ pressing, rubbing, or pinch grasp to judge the ripeness of soft fruits [9], [98], or encounter diverse levels of pleasantness under the same stroking over the arm [18]. In addition to these observed differences, individuals also inherently differ in their skin's mechanics, finger size and neural afferent density, and age-related factors tied to neural sensitivity; and thus, internal percepts of touch [17], [30].

Individuals interact in unique ways with new haptic displays [99], [100]. Likewise, the lack of tactile acuity for such haptic devices seems to arise primarily from the failure of designers to distinguish individualized characteristics [99]–[101]. Specifically, physical contact characteristics – e.g., contact regions, finger postures, surface pressure – vary widely across distinct users but are highly consistent within an individual [99]. Therefore, replicating a universal tactile paradigm alone does not afford perceptual acuity for distinct individual users, and may attenuate the fidelity and realism of tactile displays. The task of investigating the perceptual mechanism that underlies individual differences remains timely and relevant.

As the primary interface in tactile exploration, the material properties of an individual's finger pad skin may directly affect their perceptual judgments [30], [31]. Indeed, naturalistic variance in skin mechanics – e.g., surface geometry, stiffness, elasticity – can directly impact mechanotransduction in cutaneous afferents, thus, forming individuals' tactile sensitivity [27], [33], [102]. In particular, skin sites with lower thickness exhibit a lower perceptual threshold to applied pressure [28]. Meanwhile, the increased skin elasticity, or "hardness", could render a lower perceptual sensitivity to surface pressure stimulus [28], [37]. Similar to elasticity, individuals with more compliant finger pad skin exhibit substantially lower perceptual thresholds in grating orientation discrimination, especially for younger individuals [30]. Furthermore, there is gathering evidence showing that individuals' tactile acuity differs between fingertip size and gender. Specifically, improved tactile spatial acuity is coupled with decreased fingertip size [17], [103], thus, superior perception performance is found in women who on average have relatively smaller fingers than men [17], [27]. Indeed, a higher density of tactile afferents in smaller fingers may result in a finer-grained neural encoding of tactile stimuli [17], [31]. By modulating the firing properties of tactile afferents, distinct skin material properties could impact sensitivity between individuals.

In summary, it remains unclear how and if distinct skin mechanics impact perceptual performance between individuals. As a step in this direction, this work studies individuals' tactile acuity in discriminating compliances amidst inherent constraints of skin elasticity. In particular, we focus on younger participants and stimuli near the limit of tactile discriminability. By employing compliant spheres that vary in elasticity and curvature to afford non-distinct cutaneous, contact area cues, we delineate the roles of individual differences in skin mechanics and volitional exploratory movements in impacting an individual's acuity.

## 5.2 Materials and Methods

The work herein investigates how skin mechanics and exploration strategy might modulate individual differences in tactile acuity, particularly, in discriminating compliances. The methods include computational modeling of individuals' skin mechanics, as characterized by compression loading, whereas biomechanical measurements and psychophysical evaluations are conducted in differentiating small-compliant and large-stiff spheres in bare finger touch. Specifically, a finite element model of the distal finger pad was fitted for each individual to derive elastic moduli of one's skin under compression. Then, cutaneous cues evoked upon finger-to-stimulus contact – e.g., interior stress magnitude and gross contact area – were quantified in both passive and active touch interactions, via numerical simulations, customized setups, and biomechanical measurements. Finally, psychophysical responses in discriminating softness were evaluated using the psychophysical and signal detection methodologies. The derived tactile acuity for individuals was compared with corresponding skin elasticity and resultant tactile cues, thus revealing how individual differences in somatosensory perception could be attributed to distinct skin material properties and contact mechanics.

### 5.2.1 Geometry and Material Properties of the Model

The material elasticity of each participant's finger skin was characterized using two finite element models. Derived from the 3D geometry of the human distal phalanx bone [68], two simplified 2D models with plane-strain and axisymmetric elements were constructed [68]. Specifically, finger bones and nails were modeled as analytic rigid bodies and three layers of soft tissues were modeled as deformable bodies wrapped around the bone, namely epidermis, dermis, and hypodermis (Fig 5.1A). Triangular meshes were used throughout models with about 0.25 mm wide elements.

Hyperelastic material properties were used of the Neo-Hookean form of the strain energy function. The material elasticity was referred to as its shear modulus so that only one parameter is needed for material fitting, resulting in a more robust calibration. Instead of using a linear Young's modulus, a hyperelastic form was used to better simulate the deformation of soft objects in a finite-strain region.

### 5.2.2 Fitting Elasticity Ratio of Skin Layers

The initial modulus of hypodermis was set as 1 kPa [104] and the ratio search ranges for dermis and epidermis were defined [68]. Surface deflection data from *in vivo* experiments were used in the plane-strain model to fit the elasticity ratio since deflection is only controlled by the layer ratio. Specifically, an exhaustive fitting was employed: with each candidate ratio, two rigid cylindrical stimuli (diameter of 3.17 and 9.52 mm) were used to simulate indentation into the finger pad at six displacements (0.5, 1.0, 1.6, 2.5, 3.0, and 3.5 mm). The skin surface deflection was then derived and compared with empirical results from [68]. The average of all ratios with $R^2 \geq 0.8$ were derived as the final elasticity ratio of epidermal, dermal, and hypodermal layers: 510.63: 21.37: 1.00.

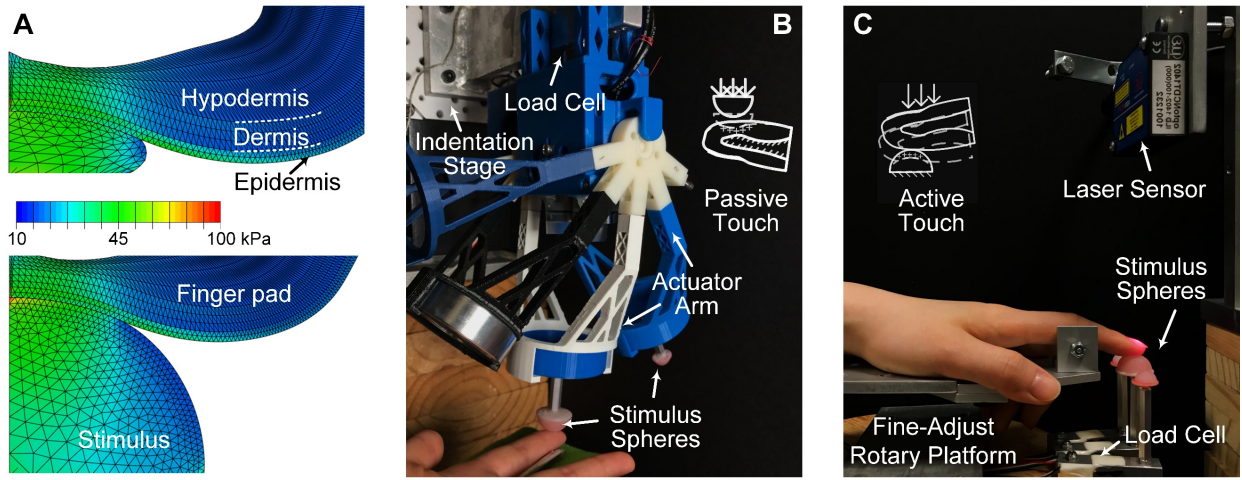

**Figure 5.1 Computational modeling and experimental platforms for the finger pad-to-stimulus contact.** (A) In finite element analysis, skin mechanics of the distal finger pad are modeled in simulated contact interactions with compliant stimuli. Hyperelastic material properties of three tissue layers are characterized and spatial distribution of stress is derived as the cutaneous response. (B) In passive touch, the spherical stimulus is indented into the stationary finger pad where the normal contact force is measured by a uniaxial load cell. (C) In active touch, the fixed stimulus is contacted by the index finger of the participant's own volition. Touch force and fingertip movement are measured by the load cell and laser sensor, respectively.

### 5.2.3 Fitting Elastic Moduli of Skin Layers

After the ratios and surface deflection were fit, elastic moduli of materials were scaled to fit the force-displacement responses measured for each participant. Similar to [33], two rigid stimuli were passively indented into the index finger pad: a flat plate and a cylinder of 6 mm diameter with a maximum 2 mm and 3 mm indentation, respectively. For each participant, by optimizing the average $R^2$ of two stimuli using the L-BFGS-B algorithm, the optimal scaling coefficient $k$ was determined. The final elastic modulus of each skin layer was then derived based on the initial shear modulus, optimal ratio, and individual scale $k$. Therefore, the reciprocal of this elasticity scale $k$ was used as a dimensionless quantity to depict the "softness" of each participant's finger pad.

### 5.2.4  Numerical Simulations

Two stimuli tips with covaried elasticity and curvature, 10 kPa-4 mm and 90 kPa-8 mm, were adopted and built [68]. Fingertip-to-stimulus contact mechanics were simulated to approximate these stimuli used in the passive touch interactions. Specifically, for each participant's fingertip model, compliant stimuli were indented into the finger pad at loads of 0.25, 0.5, 1, and 2 N. The response variable herein was derived as the cutaneous cue only, quantified as the spatial distribution of stress at the epidermal-dermal interface, where Merkel cell end-organs of slowly adapting type I afferents and Meissner corpuscles of rapidly adapting afferents reside [68].

### 5.2.5  Stimuli and Experimental Apparatus

Consistent with the finite element modeling, compliant spheres of 10 kPa-4 mm and 90 kPa-8 mm were adopted and used in empirical experiments [68]. For passive touch interactions, a customized motion stage was adopted to indent the stimulus into the stationary finger pad [68]. Customized circuitry and software were directly interfaced to control the indentation. As shown in Fig. 5.1B, stimuli were installed onto the interchangeable actuator arm and normal contact force was measured by an embedded load cell (22.2 N, 300 Hz, LCFD-5, Omega, OH). With physical constraint measures, the index finger was held at approximately 30° to the stimulus surface.

For active touch interactions, a setup based on a fine-adjust rotary platform was adopted from [9]. As shown in Fig. 5.1C, the contact force was measured by instrumented load cells (5 kg, 80Hz, TAL220B, HTC Sensor), and fingertip displacements were measured by a laser triangulation sensor (10 μm, 1.5kHz, optoNCDT 1402-100, Micro-Epsilon). The forearm and hand rested on a parallel beam with no constraints.

### 5.2.6 Measurements of Physical Contact Cues

For the fingertip displacement cue, recordings were first smoothed to remove any artifacts by a moving average filter [9]. Final displacement was derived as the absolute difference between the initiation and conclusion of the movement. Besides, to quantify contact area cues, the ink-based method was adopted to measure the gross contact area between the finger pad and stimulus for each indentation [9]. Specifically, washable ink was fully applied to the stimulus surface before each measurement. After contact, the stamped ink was transferred onto a sheet of white paper for scanning. Before each new trial, the remaining ink was removed from the finger pad. The gross contact area was derived by Gauss's formula based on the identified contact region and scaled pixels.

### 5.2.7 Data Analysis

To analyze fingertip displacements across all participants, a normalization procedure was applied since individuals had distinct ranges of finger movements. Individual displacements were normalized to the range of (0, 1) by a sigmoid membership function. The center of the transition area was the average of the data normalized and the growth rate was 1 [9].

### 5.2.8 Participants

The human-subjects experiments were approved by the Institutional Review Board at the University of Virginia. Eight naïve participants (4 females, 4 males, 28.8 ± 2.6 years of age) were recruited with written informed consent. No evidence of upper extremity pathology was reported. All participants showed right-hand dominance and completed all experimental tasks with no data was discarded.

### 5.2.9 Experimental Procedures

**Task 1 – Biomechanical measurement in passive touch.** To measure the biomechanical relationship between touch force and gross contact area in passive touch, both stimuli (10 kPa-4 mm and 90 kPa-8 mm) were indented into each participant's finger pad at 2 N load for three times. Each stimulus was ramped into the finger pad for one second and retracted for one second. The ink-based procedure was applied after each indentation for gross contact area measurements.

 **Task 2 – Biomechanical measurement in active touch.** To measure the gross contact area in active touch, both stimuli (10 kPa-4 mm and 90 kPa-8 mm) were contacted three times by each participant's index finger at 2 N load. Specifically, participants were instructed to actively press into the stimulus and a sound alarm was triggered to end the current exploration when the imposed force reached 2 N. The ink-based procedure was then applied to measure the gross contact area.

 **Task 3 – Psychophysical experiments in passive touch.** Psychophysical discrimination of compliances was conducted to evaluate individual perceptual sensitivity. Following the rule of ordered sampling with replacement, four stimulus pairs were prepared: (10,4) & (10,4), (10,4) & (90,8), (90,8) & (10,4), and (90,8) & (90,8). The test order within each pair was determined as (first) & (second). Participants were blindfolded to eliminate visual cues and no feedback was given. Using the same-different procedure, after exploring one stimulus pair (one touch per stimulus), participants reported whether the compliances of the two stimuli were the same or different. This procedure fits well with the task scope since participants can utilize any cues that are available and applicable. In passive touch, within each trial, stimuli from one pair were ramped into the finger pad successively, with an interval of 2 seconds. The indentation rate was 1 N/s and the terminal load was 2 N, as consistent with **Task 1**. For each participant, there were two trials for each stimulus pair. All trials were separated by a 15-second break and the test order was randomized to balance the response bias.

**Task 4 – Psychophysical experiments in active touch.** Discrimination tasks were conducted under participants' fully active, behavioral control. The same psychophysical procedure and stimulus pairs were employed as in **Task 3**. Within each trial, participants actively explored the compliances by contacting each stimulus (from one pair) successively, with an interval of 2 seconds. A sound alarm was triggered to end current exploration when imposed force reached 2 N load. Fingertip displacements were measured simultaneously. For each participant, there were three trials for each stimulus pair. Time break between trials was consistent with **Task 3** and test orders were also randomized to balance carry-over effects.

## 5.3  Results

### 5.3.1  Skin Elasticity Differs between Individuals

To collect force-displacement at the finger pad, uniaxial compression tests were performed (Fig 5.2A). The elasticity of an individual's skin layers was computationally characterized by curving fitting. As noted in *Methods*, by optimizing the scaling coefficient *k*, responses from each individual's model well matched their experimental measurements, with an average $R^2$ of 0.968 (Fig 5.2A). Final elastic moduli per participant were derived by the initial modulus of hypodermis (1 kPa), final elasticity ratio (510.63: 21.37: 1.00), and individual's scale *k*. Since the other two parameters are fixed, an individual's skin elasticity is solely proportional to the individual's scale *k*. Thus, the coefficient $k^{-1}$ was used to quantify the "softness" of finger pad skin, i.e., the higher $k^{-1}$ value related to the softer skin. For instance, with the optimal $k$ of 3.52 for Subject 2, the elastic moduli for epidermis, dermis, and hypodermis were derived as 1.80 MPa, 75.30 kPa, and 3.52 kPa, respectively. Each individual's skin material properties are detailed in Table 5.1 and are comparable to results by Wu et al. [104].

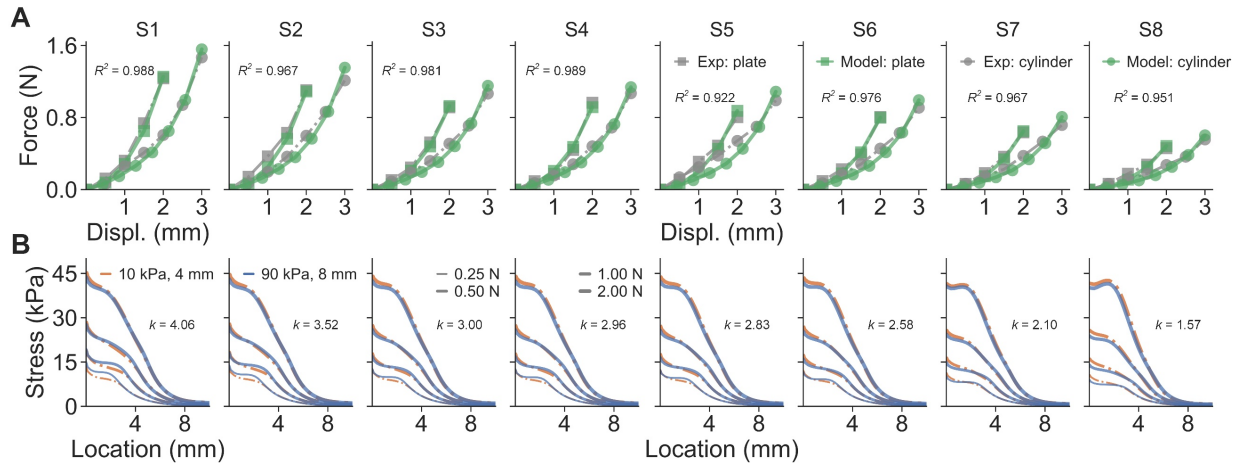

**Figure 5.2 Individual differences in skin hyperelastic material properties and stress distributions interior to the skin's layers upon passive contact.** (A) Uniaxial compression tests are performed on an individual's finger pad to derive force-displacement curves. Model generated elastic moduli are found by optimizing an individual's finger model to match experimental measurements. (B) For small-compliant (10 kPa-4 mm) and large-stiff (90 kPa-8 mm) spheres, stress distributions at the epidermal-dermal interface are nearly identical within each participant, across the four force loads. Individual differences in cutaneous responses may be attributed to skin softness, where a higher elasticity may lead to higher interior stress.

### 5.3.2   Cutaneous Contact Simulated for Individuals

To help evaluate the relationship between skin mechanics and tactile discriminability of compliance, we used the model to simulate spatial distributions of compressive stress interior to skin layers. Within each participant, stress distributions at the contact locations were nearly identical (Fig 5.2B). This aligns with prior work and indicated that only non-distinct cutaneous cues are perceptible [68].

**Table 5.1 Skin Material Properties per Individual Subject.** A higher $k^{-1}$ value is related to the softer skin.

| Sub. | $k^{-1}$ * | Epidermis (MPa) | Dermis (kPa) | Hypodermis (kPa) |
|---|---|---|---|---|
| S1 | 0.25 | 2.07 | 86.66 | 4.06 |
| S2 | 0.28 | 1.80 | 75.30 | 3.52 |
| S3 | 0.33 | 1.53 | 64.14 | 3.00 |
| S4 | 0.34 | 1.51 | 63.28 | 2.96 |
| S5 | 0.35 | 1.44 | 60.45 | 2.83 |
| S6 | 0.39 | 1.32 | 55.22 | 2.58 |
| S7 | 0.48 | 1.07 | 44.84 | 2.10 |
| S8 | 0.64 | 0.80 | 33.52 | 1.57 |

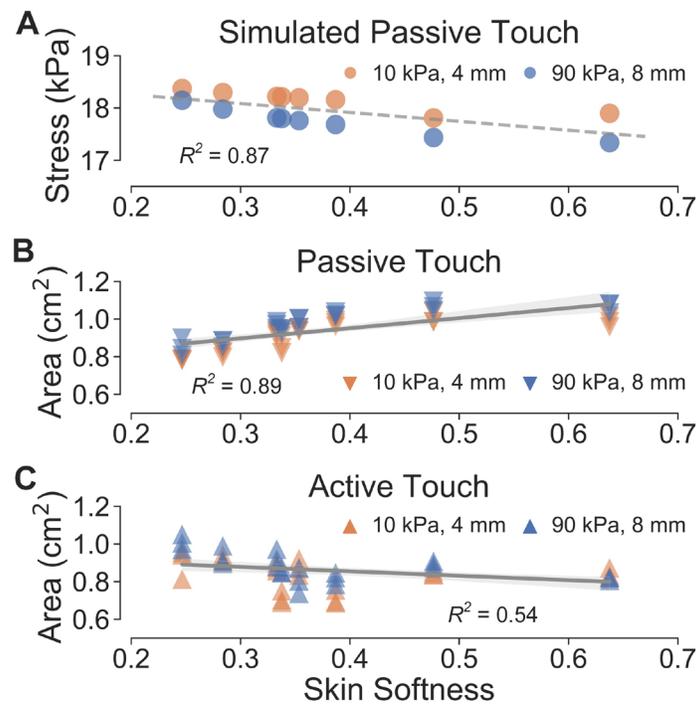

**Figure 5.3 Biomechanical relationships between individuals' skin softness ($k^{-1}$) and their cutaneous contact cues** represented by (A) simulated average interior stress in passive touch, and gross contact areas measured at 2N in (B) passive and (C) active touch. Each data point denotes one measurement. Linear regression is applied for each cutaneous cue and the goodness of fit is indicated by $R^2$. In passive touch under the same contact load, individuals with softer skin proportionally exhibited lower (A) interior stress and larger (B) gross contact areas. Overall, within an individual, non-distinct cutaneous cues are consistently elicited by the two spherical stimuli. Therefore, individual differences in discrimination might be impacted by the magnitude of afferents recruitment, originally derived from the skin softness.

To further investigate individual differences in cutaneous contact, average interior stress (within lateral contact locations from 0 to 10 mm, at 2 N load) was calculated per stimulus and participant. As shown in Fig 5.3A, for the same indentation magnitude and rate, greater interior stress is observed for those individuals with stiffer skin. The linear regression yielded a strong negative correlation between skin softness and average stress of two stimuli, with a Spearman's coefficient of -0.99. This model result illustrates that differences in individual skin mechanics may indeed evoke variance in interior stress magnitude, which thus, could potentially lead to individual differences in perceptual acuity.

### 5.3.3 Individual Differences in Contact Biomechanics

Derived from the finite element simulation, we computationally showed that skin mechanics may evoke individual differences in contact responses. To further validate this observation, gross contact area was measured in passive and active touch interactions. In passive touch within an individual, gross contact areas for these two stimuli indeed overlapped (Fig 5.3B). Moreover, the linear regression yielded a strong positive correlation between individuals' skin softness ($k^{-1}$) and average gross contact area with Spearman's coefficient of 0.87 ($p$ = 2.01e-15). This illustrates that a larger gross contact area is evoked by individuals with softer skin.

In active touch within an individual, gross contact areas of two stimuli remained overlapped (Fig 5.3C), as consistent with the passive touch (Fig 5.3B) and computational modeling (Fig 5.2B) results. Furthermore, the linear regression yielded a negative correlation between individuals' skin softness and average gross contact area with Spearman's coefficient of -0.58 ($p$ = 1.38e-5). However, opposite to the results in passive touch, larger gross contact areas were evoked by individuals with harder skin. This indicates that, in active touch, individuals could move their fingers differently to evoke additional cutaneous cues as opposed to passive touch.

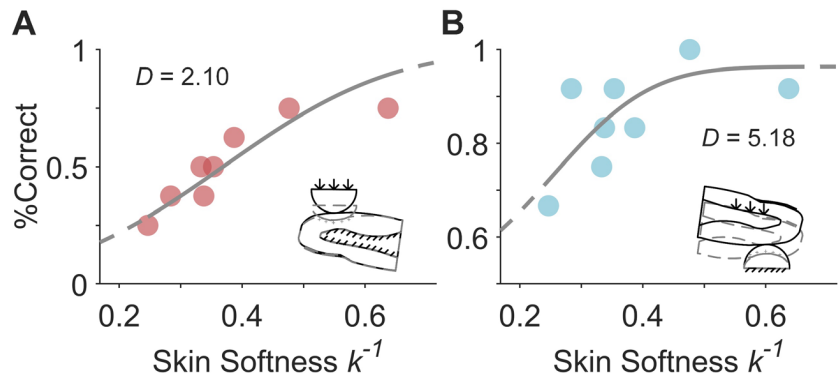

**Figure 5.4 Relationships between skin softness and discrimination performance are characterized by the psychometric curve.** The goodness of fit is indicated by deviance. (A) In passive touch, with only cutaneous cues available, individual correctness is proportional to skin softness. (B) In active touch, with additional perceptual cues, individual performance is improved but with a weaker correlation between skin softness.

### 5.3.4 Individual Differences in Perceptual Sensitivity

To further investigate whether individuals' skin mechanics could impact perceptual performance, relationships between individuals' skin softness and psychophysical responses were analyzed. Specifically, for each participant, the aggregated percentage of correct discriminations for all stimulus pairs was calculated and compared with the individual's skin softness. The correlation was quantified with the psychometric function, which was fitted by the beta-binomial model and the goodness of fit was indicated by the deviance [105]. In passive touch (Fig 5.4A), discrimination correctness was positively correlated with the skin softness, where individuals with softer fingers indeed achieved better performance. For all participants aggregated, the average percentage of correctness was 51.6% ± 18.2, which is indeed at the limit of discriminability. In active touch (Fig 5.4B), individual's performance was all improved but the correlation was weaker than passive touch ($D = 5.18$). For all participants aggregated, the average correctness was indeed improved to 85.4% ± 10.7.

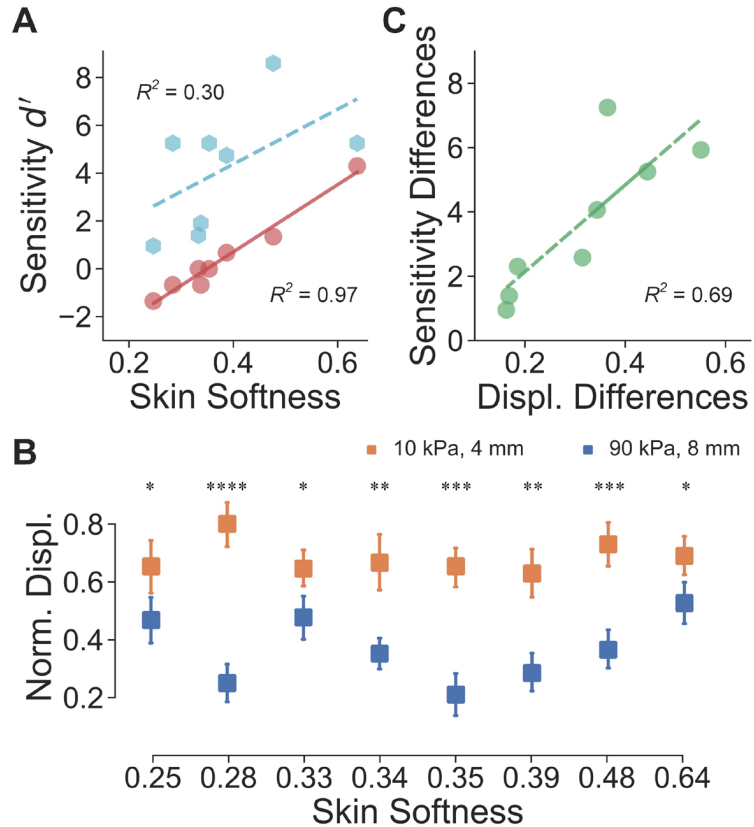

**Figure 5.5 Relationships between individual skin softness, perceptual sensitivity, and exploratory strategies in active touch.** (A) In passive touch (red) where cutaneous cues are non-distinct, individual perceptual sensitivity could be derived by skin softness. In active touch (blue) where supplementary cues are available, individual sensitivity is improved. (B) In active touch, participants move their fingers to evoke significant differences in displacements for discrimination. (C) The improvement in individual sensitivity is positively correlated with the displacement differences one evokes to help discrimination. The goodness of fit is indicated by $R^2$. Error bars denote 90% confidence intervals and *$p < 0.05$, **$p < 0.01$, ***$p < 0.001$, and ****$p < 0.0001$, by the Mann-Whitney U test.

Furthermore, for each participant, the sensitivity index $d'$ was computed to provide a bias-free measure of individual discriminability of compliance [19]. As shown in Fig 5.5A for passive touch, the linear regression yielded a strong positive correlation between skin softness and individual sensitivity index with Spearman's coefficient of 0.96 ($p$ = 1.78e-4). This indicates that, in passively discriminating compliances with non-distinct cutaneous cues, individual perceptual sensitivity could be attributed to individual skin mechanics, where participants with softer finger pads could proportionally afford higher sensitivity. Note that negative *d'* values were derived as the hit rate was lower than the false alarm rate in the passive discrimination task. This also aligns with the chance performance shown in Fig 5.4A.

However, in active touch, the correlation between skin softness and individual sensitivity became weaker (Spearman's coefficient of 0.66, *p* = 0.08). This indicates that distinct from passive touch, the improved individual sensitivity cannot be solely attributed to the skin softness. Indeed, participants actively evoke significant differences in fingertip displacements to help discriminate between these two stimuli (Fig 5.5B). Moreover, there is a positive correlation between differences in individual's displacement and individual's improvement in sensitivity from passive to active touch (Spearman's coefficient of 0.93, *p* = 8.63e-4, Fig 5.5C). This indicates that, in active touch, individuals employ distinct strategies to evoke supplementary cues, thus, optimally improving sensitivity.

## 5.4 Discussion

This work considers how individual differences in the skin's mechanics might impact perceptual acuity with compliant materials, near the limit of discriminability. By employing elastic spheres that vary in curvature and elasticity to afford non-distinct cutaneous cues, we find that individuals tend to actively control their exploratory movements to optimally improve their acuity, which is inherent, in at least this younger cohort of participants, derived by elastic material properties of their skin. That is, while those with harder skin appear to inherently perform worse in passive touch, participants can and do improve by actively changing their exploration strategies. In so doing, they move distinctly to generate contact cues (Fig 5.3C and 5.5B), which are independent of skin mechanics, and in line with prior work [68].

Therefore, there are several implications in the design of engineered haptic systems to afford individual differences. For touch-enabled displays grounded to the user's finger, we find individuals exhibit distinct contact profiles in affording surface pressure and retaining gross contact regions (Fig 3.3). These align with observations in interacting with touch devices where diverse pressure inputs [101] and contact locations [100], [106] are behaviorally applied by users. These together argue that individualized design and actuation mechanisms may render a better tactile acuity to individual users [106].

As noted, in passive touch where cutaneous cues only are perceptible, individual differences in perceptual acuity could be attributed to differences in skin elasticity. As in line with discriminating surface pressure [28], [37], and grating orientation [30], individuals with softer fingertips proportionally exhibit higher perceptual sensitivity (Fig 5.4A and 5.5A). Indeed, when stimuli are passively indented into softer finger pads, larger gross contact areas and slightly more differences between two stimuli are retained from larger surface deformations (Fig 5.3B), thus, it may well be the case that more populations of mechanoreceptors are solicited and recruited to augment individuals' tactile perception [27]. Noteworthy, as we only consider a younger cohort herein, the relation between skin mechanics and tactile acuity might be distinct for older individuals, where the general decline in acuity could be attributed to the lower mechanoreceptive sensibility [30], [31].

Furthermore, in active touch, individuals could improve their performance by individualizing their exploration strategies. In particular, to compensate for poor perceptual acuity constrained by skin elasticity, individuals with harder skin behaviorally move their fingertips to create larger gross contact areas (Fig 5.3C), which may recruit a larger number of tactile afferents. Indeed, improved tactile acuity is positively correlated with differences in proprioceptive cues employed for discrimination (Fig 5.5C). Such exploration strategy of behavioral control aligns with how we explore naturalistic soft objects [6], [9]: individuals utilize distinct fingertip displacements to readily differentiate compliances. Overall, in the context of natural exploration, individuals could actively control their movements to achieve optimal perceptual performance, even amidst the inherent constraints of skin mechanics.

Moreover, the study herein only considers elastic moduli of finger pad layers to characterize skin mechanics. Prior studies suggest that variance in skin thickness and stiffness may also affect perceptual sensitivity [27], [28]. For instance, higher skin thickness can increase distances between mechanoreceptors and surface pressure, thus, may attenuating afferent firing for perceptual threshold [28]. Ultimately, the ability to transform a surface stimulus into an afferent population response is vital to creating percepts [17], [28]. Thus, changes in skin mechanics cannot fully account for variance in individuals' tactile acuity. Factors including sex differences in density of mechanoreceptors [17], the loss of mechanoreceptive sensitivity, and changes in skin's mechanical properties with aging [30], or more behaviorally, individual exploratory movements (Fig 5.5C) may together play a role in processing stages that underlie perception.

## 5.5  Acknowledgment

This work was supported in part by grants from the National Science Foundation (IIS-1908115) and the National Institutes of Health (NINDS R01NS105241). The content is solely the responsibility of the authors and does not necessarily represent the official views of the NSF or NIH.

# 6  Overall Conclusion and Future Work

In this dissertation, we performed significant groundwork towards understanding the human tactile perception of material softness. The overall objective of this study was to decipher optimal touch contact interactions and perceptual strategies that underlie softness perception within and among individuals. We seek to address this by employing an integrated methodology of computational finite element modeling, biomechanical experimentation, and psychophysical studies. First, by developing an elasticity-curvature illusion phenomenon, we dissociated relative contributions from cutaneous and kinesthetic cues in encoding material softness. We revealed that our perception of softness is a product of both sensation and volition, and depends upon both afferents in skin and proprioception. Second, in exploring both engineered and ecological soft objects, we found that similar exploratory strategies are volitionally employed among individuals. Indeed, touch force and finger movements are finely tuned such that optimal sensory cues and perceptual rules could be evoked for discrimination amidst movement control, material, and exploration time. Third, considering inherent differences among individuals' skin and behavior, we investigated individual differences in skin mechanics, exploratory strategies, and thus, perceptual sensitivity. We revealed that an individual's tactile acuity is essentially constrained by inherent skin material properties, but could be improved under volitional control of exploration movements.

By studying the elasticity-curvature illusion, we've learned that pressing an object into the stationary finger does not reveal its softness, but pressing actively does. This phenomenon illuminates an interplay within our somatosensory system, in particular, between cutaneous responses from skin receptors and proprioceptive feedback traditionally tied to joint movements. It also reveals how our movements optimally evoke these cues to inform our perception of softness. Indeed, across a range of touch interactions broader than just softness, we find that cutaneous and proprioceptive cues are integrated to achieve high levels of performance [41], [62]. In tasks involving reaching movements, cutaneous cues could systematically bias motion estimates, indicating that multisensory cues are optimally integrated for our motor control [41]. In general, multimodal interactions between these two signals are found to be mediated by the distinct neural mechanism in the primary somatosensory cortex [62]. These findings come in general agreement with prior studies reporting that both cutaneous and proprioceptive cues are needed in discriminating compliance. In particular, when finger movements are eliminated, our ability to discriminate pairs of spring cells decreases [7]. Likewise, when pinching an elastic substrate in-between two rigid plates, lower discriminability of compliance is obtained when relying upon proprioception alone as compared to cutaneous cues alone [5].

Our work to analyze contact interactions with the fruit – the first of its kind to the authors' knowledge – quantifies touch interaction cues and exploration strategies with a natural object, in particular, the palpation of soft plum fruit. In nearly all prior studies, however, stimuli have been highly engineered and delivered by sophisticated devices [46], like silicone-elastomers and foams. To enable study with ecological substances, a novel experimental paradigm was developed to measure material properties and touch contact of fruit, which can break down rapidly between sessions. Meanwhile, several aspects regarding the experimental design could be taken into account for future work due to inherent difficulties in working with delicate objects that change over time. To be more definitive, a more robust experimental design will be required, of a nature distinct from the typical two-alternative forced-choice variety, due to inherent difficulties in working with easily damageable objects that change over time, such as ripe fruit. As noted in *4.2.7*, after several trials with three participants, one of the plums had reached a state where the experimenters felt – though rotated to distinct touch points between participants – it was irrevocably damaged and would not yield reliable results. Understanding when damage has gone beyond a threshold – let alone avoid damage – may require solutions at conflict with experimental sample size requirements. Regarding the latter, one would want to use many participants, have participants touch in spots that are similar in firmness, and have multiple trials per subject. However, with soft fruit, meeting all of these requirements is a challenge. Moreover, the differences between the two plums used herein were fairly obvious, as indicated in Fig 4.1C. One would anticipate wanting to evaluate a greater variety of specimens, with some at the limit of discriminability for the best performing participants. There are timing issues as well to take into account. We acquired specimens and did the experiments on the same day with two of the participants with the third participant completing the experiment on the morning of the second day. There are means such as control of temperature, humidity, and light with which to account, should an experimental paradigm require a longer duration.

As noted, there is a consensus on the role of peak force as a reliable perceptual cue in deciphering softness [6], [8], [23], [24]. In particular, prior studies indicate that higher force and force-rate are applied when exploring harder stimuli as opposed to softer ones [6], [8], [23], [65]. The peak indentation force may be tuned to achieve optimal performance amidst impact from stimuli differences [23], prediction to exploration targets [24], prior experiences on similar stimuli [107], and rewarding mechanisms [24]. Furthermore, the integration of touch force and finger displacement indeed encode perception of compliance and could be utilized for psychophysical discrimination [5]–[8], [23], [65]. Such correlation can be quantified by the virtual stiffness cue which affords reliable estimates of stiffness perception [65], [83]. For instance, the maximum force and corresponding finger displacement at the end of compression were employed for the compliance judgment [65]. The change rate of the force-displacement curve was also considered as an efficient temporal cue which was quantified by regression slopes [108]. However, within one single exploration, the recognition of compliance could be modeled as an optimization procedure that recursively finds a stiffness estimate that best fits the perceived tactile cues [83]. This procedure keeps updating the estimate to the true value by the fusion of historical and current inputs. Indeed, gathering evidence indicates that exploration duration and repeat time could impact the integration of tactile information [21], [109], [110], which could be considered as one of the future directions.

# Publications

## Papers in Progress

[P1]  **C. Xu**, Y. Wang, G. Gerling, Perceptual strategies are fine-tuned to elicit optimal mechanosensory cues in tactile decision making, in writing, to be submitted to *J. Royal Society Interface*, 2022.

[P2]  S. Xu, **C. Xu**, S. McIntyre, H. Olausson, G. Gerling, Subtle contact nuances in the delivery of human-to-human touch distinguish emotional sentiment, under revision, *IEEE Trans. Haptics*, 2021.

[P3]  A. Kao, **C. Xu**, G. Gerling, Using digital image correlation to quantify skin deformation with von Frey monofilaments, under revision, *IEEE Trans. Haptics*, 2021.

[P4]  S. Xu, **C. Xu**, S. McIntyre, H. Olausson, G. Gerling, 3D visual tracking to quantify physical contact interactions in human-to-human touch, in writing, to be submitted to *Front. Physiol.*, 2021.

[P5]  N. Yu, **C. Xu**, An improved wrist kinematic model for human-robot interaction, *arXiv*, 2020.

## Journal Articles

[J1]  **C. Xu**, Y. Wang, G. Gerling, An elasticity-curvature illusion decouples cutaneous and proprioceptive cues in active exploration of soft objects, *PLOS Comput. Biol.*, 17(3): e1008848, 2021. (IF 4.475)

[J2]  **C. Xu**, H. He, S. Hauser, G. Gerling, Tactile exploration strategies with natural compliant objects elicit virtual stiffness cues, *IEEE Trans. Haptics*, 13(1): 4-10, 2020. (IF 2.487)

[J3]  N. Yu, S. Wang, **C. Xu**, RGB-D based autonomous exploration and mapping of a mobile robot in unknown indoor environment, *Robot*, 39(6): 860-871, 2017. (in Chinese)

[J4]  N. Yu, S. Li, Y. Zhao, K. Wang, **C. Xu**, Design and implementation of a dexterous human-robot interaction system, *Chinese Journal of Scientific Instrument*, 36(3): 602-611, 2017. (in Chinese)

[J5]  N. Yu, **C. Xu**, H. Li, K. Wang, L. Wang, J. Liu, Fusion of haptic and gesture sensors for rehabilitation of bimanual coordination and dexterous manipulation, *Sensors*, 16(3): 395, 2016. (IF 3.576)

[J6]  N. Yu, Y. Li, **C. Xu**, J. Liu, Localization and motion planning and control of mobile robots with RGB-D SLAM, *Journal of Systems Science and Mathematical Sciences*, 35(7): 838-847, 2015. (in Chinese)

## Peer-Reviewed Conference Publications

[C15] N. Yu, **C. Xu,** K. Wang, Z. Yang, J. Liu, Gesture-based telemanipulation of a humanoid robot for home service tasks, *IEEE Int. Conf. CYBER Technol. Autom. Control Intell. Syst. (CYBER),* 2015.

## Peer-Reviewed Conference Abstract

[A1]  **C. Xu**, H. He, S. Hauser, G. Gerling, Measurement of touch interaction cues in discriminating soft fruit, *IEEE World Haptics Conference (WHC),* 2019.
91

1991.

[70] K. Dandekar, "Role of mechanics in tactile sensing of shape," Massachusetts Institute of Technology, 1995.

[71] R. J. Gulati, M. A. Srinivasan, and others, "Human fingerpad under indentation I: static and dynamic force response," *ASME-Publications-Bed*, vol. 29, p. 261, 1995.

[72] G. A. Gescheider, *Psychophysics: The Fundamentals*, Third Edit. Lawrence Erlbaum Associates Publishers, 1997.

[73] W. A. Wagenaar, "Sequential response bias in psychophysical experiments," *Percept. Psychophys. 1968 35*, vol. 3, no. 5, pp. 364–366, Sep. 1968.

[74] A. C. Zoeller and K. Drewing, "A Systematic Comparison of Perceptual Performance in Softness Discrimination with Different Fingers," *Attention, Perception, Psychophys.*, pp. 1–14, Jul. 2020.

[75] X. D. Pang, H. Z. Tan, and N. I. Durlach, "Manual discrimination of force using active finger motion," *Percept. Psychophys.*, vol. 49, no. 6, pp. 531–540, Nov. 1991.

[76] L. K. Klein, G. Maiello, V. C. Paulun, and R. W. Fleming, "Predicting precision grip grasp locations on three-dimensional objects," *PLOS Comput. Biol.*, vol. 16, no. 8, p. e1008081, Aug. 2020.

[77] G. Westling and R. S. Johansson, "Factors influencing the force control during precision grip," *Exp. Brain Res.*, vol. 53, no. 2, pp. 277–284, Jan. 1984.

[78] B. B. Edin and N. Johansson, "Skin strain patterns provide kinaesthetic information to the human central nervous system," *J. Physiol.*, vol. 487, no. 1, pp. 243–251, Aug. 1995.

[79] A. Moscatelli *et al.*, "A change in the fingertip contact area induces an illusory displacement of the finger," in *Lecture Notes in Computer Science (including subseries Lecture Notes in Artificial Intelligence and Lecture Notes in Bioinformatics)*, 2014, vol. 8619, pp. 72–79.

[80] A. V. Terekhov and V. Hayward, "The brain uses extrasomatic information to estimate limb displacement," *Proc. R. Soc. B Biol. Sci.*, vol. 282, no. 1814, p. 20151661, Sep. 2015.

[81] S. C. Hauser, S. S. Nagi, S. McIntyre, A. Israr, H. Olausson, and G. J. Gerling, "From Human-to-Human Touch to Peripheral Nerve Responses," in *2019 IEEE World Haptics Conference (WHC)*, 2019, pp. 592–597.

[82] S. C. Hauser, S. McIntyre, A. Israr, H. Olausson, and G. J. Gerling, "Uncovering Human-to-Human Physical Interactions that Underlie Emotional and Affective Touch Communication," in *IEEE World Haptics Conference (WHC)*, 2019, pp. 407–412.

[83] B. Wu and R. L. Klatzky, "A recursive bayesian updating model of haptic stiffness perception," *J. Exp. Psychol. Hum. Percept. Perform.*, vol. 44, no. 6, pp. 941–952, 2018.

[84] M. Kuschel, M. Di Luca, M. Buss, and R. L. Klatzky, "Combination and Integration in the Perception of Visual-Haptic Compliance Information," *IEEE Trans. Haptics*, vol. 3, no. 4, pp. 234–244, Oct. 2010.

[85] N. Gurari, K. J. Kuchenbecker, and A. M. Okamura, "Perception of springs with visual and proprioceptive motion cues: Implications for prosthetics," *IEEE Trans. Human-Machine Syst.*, vol. 43, no. 1, pp. 102–114, Jan. 2013.

[86] M. D. Rinderknecht *et al.*, "Reliability, validity, and clinical feasibility of a rapid and objective assessment of post-stroke deficits in hand proprioception," *J. Neuroeng. Rehabil.*, vol. 15, no. 1, p. 47, Dec. 2018.

[87] D. Katz, "The judgements of test bakers. A psychological study," *Occup. Psychol.*, vol. 12, pp. 139–